\documentclass[aps,twocolumn,showpacs,citeautoscript,reprint,prl]{revtex4-2}

\usepackage{bm}
\usepackage{graphicx}
\usepackage{color}
\usepackage[binary,amssymb]{SIunits}
\usepackage{comment}

\usepackage{amsmath}

\usepackage{color}
\definecolor{Blue}{rgb}{0.3,0.3,0.9}
\definecolor{Red}{rgb}{0.9,0.3,0.3}
\definecolor{Green}{rgb}{0.3,0.6,0.3}

\begin{document}

\begin{abstract}

Electron hydrodynamics encompasses the exotic fluid-like behavior of electrons in two-dimensional materials such as graphene. It accounts for superballistic conduction, also known as the Gurzhi effect, where increasing temperature reduces the electrical resistance. In analogy with conventional fluids, the Gurzhi effect is only expected in the hydrodynamic regime, with the decrease in the resistance occurring at intermediate temperatures. Nonetheless, experiments on electron fluids consistently show that superballistic conduction starts at close-to-zero temperatures. To address this paradox, we study hydrodynamic flow, and we find that replacing the classical dynamics with tomographic dynamics, where only head-on collisions are allowed between electrons, solves the dilemma. The latter strengthens superballistic conduction, with potential applications in low-dissipation devices, and explains its differences with the Molenkamp effect and conventional fluids dynamics. Our study reveals that the superballistic paradox is resolved by considering the electrons not as classical particles but as fermions.  
 
\end{abstract}

\title{Superballistic paradox in electron fluids:
Evidence of tomographic transport}

\author{Jorge Estrada-\'{A}lvarez}

\affiliation{GISC, Departamento de F\'{\i}sica de Materiales, Universidad 
Complutense, E-28040 Madrid, Spain}

\author{Elena D\'{i}az}

\affiliation{GISC, Departamento de F\'{\i}sica de Materiales, Universidad 
Complutense, E-28040 Madrid, Spain}

\author{Francisco Dom\'{i}nguez-Adame}

\affiliation{GISC, Departamento de F\'{\i}sica de Materiales, Universidad 
Complutense, E-28040 Madrid, Spain}

\date{\today}

\maketitle

The pursuit of miniaturization of electronic devices faces the inherent challenge of energy dissipation~\cite{charge_transport_and_hydrodynamics_in_materials}, and further optimizing the devices relies on our ability to mitigate the increased electrical resistance. A feasible strategy to achieve this goal is to use the Gurzhi effect, exploiting electrons' hydrodynamic behavior~\cite{hydrodynamic_approach_to_two_dimensional_electron_systems,viscous_electron_fluids,a_perspective_on_non_local_electronic_transport_in_metals_viscous_ballistic_and_beyond}. Electron hydrodynamics substitutes ohmic transport in many two-dimensional~(2D) materials, such as graphene and gallium arsenide heterostructures~\cite{difussion_of_photo_excited_holes_in_viscous_electron_fluid,negative_magnetoresistance_in_viscous_flow_of_two_dimensional_electrons,temperature_dependence_of_electron_viscosity_in_superballistic_gaas_point_contact,geometric_control_of_universal_hydodynamic_flow_in_a_two_dimensional_electron_fluid,nonlinear_transport_phenomena_and_current_induced_hydrodynamics_in_ultrahigh_mobility_two_dimensional_electron_gas,hydrodynamic_and_ballistic_transport_over_large_length_scales_in_gaas_algaas}, or PdCoO$_2$~\cite{evidence_for_hydrodynamics_electron_flow_inPdCoO}. Electron fluids exhibit exotic signatures~\cite{observation_of_electronic_viscous_dissipation_in_graphene_magneto_thermal_transport,breaking_down_the_magnonic_wiedemann_franz_law_in_the_hydrodynamic_regime}, from Poiseuille's flow~\cite{visualizing_poiseuille_flow_of_hydrodynamic_electrons,imaging_the_breakdown_of_ohmic_transport_in_graphene} to the formation of whirlpools~\cite{direct_observation_of_vortices_in_an_electron_fluid,observation_of_current_whirlpools_in_graphene_at_room_temperature,negative_local_resistance_caused_by_viscous_electron_backflow_in_graphene,linking_spatial_distributions_of_potential_and_current_in_viscous_electronics}, and they have potential applications in 2D devices, including high-frequency operation~\cite{viscous_terahertz_photoconductivity_of_hydrodynamic_electrons_in_graphene,negative_differential_resistance_of_viscous_electron_flow_in_graphene} and the Gurzhi effect~\cite{minimum_of_resistance_in_impurity_free_conductors,hydrodynamic_effects_in_solids_at_low_temperature}. This effect exploits the collective motion of electrons to evade scattering against the device's edges. Therefore, the resistance is lower than the ballistic limit, resulting in superballistic conduction and enabling devices with reduced dissipation~\cite{eliminating_the_channel_resistance_in_two_dimensional_systems_using_viscous_charge_flow,how_electron_hydrodynamics_can_eliminate_the_landauer_sharvin_resistance,superballistic_flow_of_viscous_electron_fluid_through_graphene_constrictions,higher_than_ballistic_conduction_of_viscous_electron_flows,optimal_geometries_for_low_resistance_viscous_electron_flow}. 

\begin{figure}[ht]
    \centering
    \includegraphics[width=1\linewidth]{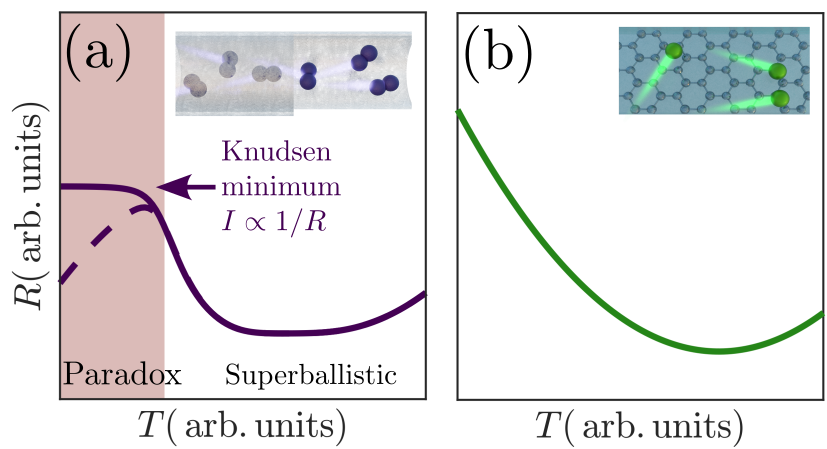}
    \caption{Superballistic paradox scheme. (a)~Inspired by conventional fluids, Gurzhi suggested a decrease in the electrical resistance of metals with increasing temperatures. The latter would only occur above a finite threshold temperature, where the  Knudsen minimum arises~\cite{minimum_of_resistance_in_impurity_free_conductors}.     
    %~Inspired by conventional fluids, which show the Knudsen minimum, Gurzhi suggested a decrease in the electrical resistance of metals that would only take place above a finite threshold temperature~\cite{minimum_of_resistance_in_impurity_free_conductors}. 
    (b)~On the contrary, experiments with electron fluids show that the decrease starts at close-to-zero temperature~\cite{boundary_mediated_electron_electron_interactions_in_quantum_point_contacts,temperature_dependence_of_electron_viscosity_in_superballistic_gaas_point_contact,superballistic_electron_flow_through_a_point_contact_in_a_gaalas_heterostructure,superballistic_flow_of_viscous_electron_fluid_through_graphene_constrictions,superballistic_conduction_in_hydrodynamic_antidot_graphene_superlattices,boundary_mediated_electron_electron_interactions_in_quantum_point_contacts,effects_of_electron_electron_scattering_in_wide_ballistic_microcontacts,long_distance_electron_electron_scattering_detected_with_point_contacts,superballistic_electron_flow_through_a_point_contact_in_a_gaalas_heterostructure,quantitative_measurement_of_viscosity_in_two_dimensional_electron_fluids,geometric_control_of_universal_hydodynamic_flow_in_a_two_dimensional_electron_fluid}.}
    \label{fig:Qualitative1}
\end{figure}

The Gurzhi effect involves a decreasing electrical resistance at intermediate temperatures, as depicted schematically in Fig.~\ref{fig:Qualitative1}(a). The flow is mostly ballistic at low temperatures, but increasing temperature favors electron-electron collisions and leads to viscous electron flow. Thus, the Gurzhi effect would only occur at intermediate temperatures~\cite{minimum_of_resistance_in_impurity_free_conductors}, once the distance that electrons travel before colliding with other electrons is shorter than the size of the device, namely $l_{ee} < d$. More precisely, as Gurzhi later suggested for 2D systems~\cite{electron_electron_collisions_and_a_new_hydrodynamic_effect_in_two_dimensional_electron_gas}, when $l_{ee} \sqrt{T/T_F} < d$, where $T_F$ is the Fermi temperature. Since $l_{ee} \sim 1/T^2$~\cite{viscous_electron_fluids}, both conditions involve that the Gurzhi effect is expected to occur above a finite threshold temperature. 
Except for collisions with phonons, which raise the resistance at even higher temperatures, most of the Gurzhi effect would resemble the behavior of a conventional fluid: at low collision rates, the resistance increases with collisions until it reaches a maximum or, equivalently, a minimum current, known as the Knudsen minimum in conventional fluids~\cite{knudsen_minimum_original,predicting_the_knudsen_paradox_in_long_capillaries_by_decomposing_the_flow_into_ballistic_and_collision_parts,predicting_the_knudsen_paradox_in_long_capillaries_by_decomposing_the_flow_into_ballistic_and_collision_parts}. 

Progress in 2D materials has enabled the realization of the superballistic effect predicted by Gurzhi. However, experiments performed in graphene and gallium arsenide heterostructures
%, which explored several geometries, 
show a different behavior: The decrease in the resistance starts at close-to-zero temperatures~\cite{temperature_dependence_of_electron_viscosity_in_superballistic_gaas_point_contact,superballistic_electron_flow_through_a_point_contact_in_a_gaalas_heterostructure,superballistic_flow_of_viscous_electron_fluid_through_graphene_constrictions,superballistic_conduction_in_hydrodynamic_antidot_graphene_superlattices,boundary_mediated_electron_electron_interactions_in_quantum_point_contacts,effects_of_electron_electron_scattering_in_wide_ballistic_microcontacts,long_distance_electron_electron_scattering_detected_with_point_contacts,superballistic_electron_flow_through_a_point_contact_in_a_gaalas_heterostructure,quantitative_measurement_of_viscosity_in_two_dimensional_electron_fluids}, as shown schematically in Fig.~\ref{fig:Qualitative1}(b), and no signatures of the Knudsen minimum are observed. Thus, experimental evidence supports the prediction of resistance reduction due to electron-electron collisions,. Still, such reduction is already effective at temperatures much lower than those estimated by Gurzhi.

\begin{figure*}[ht]
    \centering
    \includegraphics[width=1\linewidth]{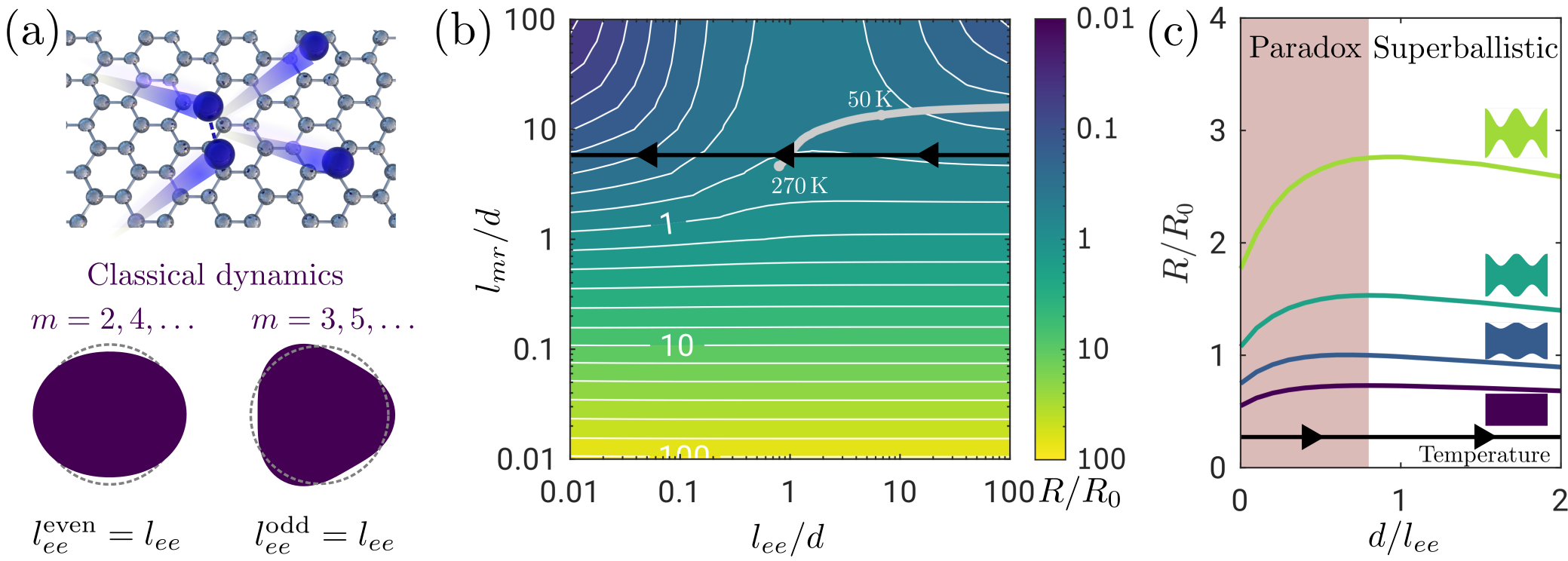}
    \caption{Classical dynamics. (a)~Electrons can collide regardless of their direction of movement, allowing for the relaxation of the even and odd parity modes. (b)~ resistance of a uniform channel as a function of the collision rates, $l_{ee}$ and $l_{mr}$. The gray line shows typical values for a graphene channel of width $d = 200 \, \rm nm$ at $n=0.5\times 10^{12} \, \rm cm^{-2}$~\cite{superballistic_flow_of_viscous_electron_fluid_through_graphene_constrictions,negative_local_resistance_caused_by_viscous_electron_backflow_in_graphene}. (c) Resistance as a function of $l_{ee}^{-1}\sim T^{2}$. At low temperatures, the resistance first increases with temperature, but after a threshold, it decreases. Therefore, classical dynamics predict a significant increase in the resistance for $ l_{ee} \gtrsim d$, not observed in the experiments.     
    We show the results for uniform and crenelated channels with increasing corrugation at $l_{mr} / d = 5$. }
    \label{fig:hydrodynamic}
\end{figure*}

In this Letter, we investigate this \emph{superballistic paradox} and sort it out by carefully considering the electron microscopic dynamics. The unexpected superballistic effect observed in experiments supports the tomographic description~\cite{tomographic_dynamics_and_scale_dependent_viscosity_in_2D_electron_systems} at low temperatures, and it affects the fundamentals of the Gurzhi effect. \\ 

% \textit{Fundamentals.}

Let us consider the polar distribution function $g({\bm r}, \theta)$ that accounts for the excess of electrons above the Fermi distribution at position $\bm r$ moving in the direction of $\theta$. Since transport phenomena affect electrons near the Fermi line, $g({\bm r}, \theta)$ satisfies the following Boltzmann transport equation~\cite{ballistic_hydrodynamic_phase_transition_in_flow_of_two_dimensional_electrons,magnetic_field_suppression_of_tomographic_electron_transport,effects_of_electron_electron_scattering_in_wide_ballistic_microcontacts,ballistic_and_hydrodynamic_magnetotransport_in_narrow_channels,alternative_routes_to_electron_hydrodynamics,hydrodynamic_magnetotransport_in_two_dimensional_electron_systems_with_macroscopic_obstacles,ballistic_flow_of_two_dimensional_interacting_electrons,supp}
\begin{equation}
\left(\begin{matrix}
\cos \theta \\ \sin \theta
\end{matrix} \right) 
\cdot \nabla_{{\bm r}} \left( g- \frac{e V}{ \hbar k_F} \right)  = -\frac{g}{l_{mr}} + \Gamma_{ee} \left[ g \right] \ ,
\label{BTE}
\end{equation}  
where $k_F$ is the Fermi wavenumber, $V(\bm r)$ is the electric potential, and, according to experiments, a constant carrier density $n$ is set~\cite{superballistic_flow_of_viscous_electron_fluid_through_graphene_constrictions,superballistic_conduction_in_hydrodynamic_antidot_graphene_superlattices,geometric_control_of_universal_hydodynamic_flow_in_a_two_dimensional_electron_fluid}. It is worth mentioning that the Boltzmann transport equation does not only enable to describe the hydrodynamic regime but also the ballistic one. We 
%consider the Callaway's ansatz~\cite{model_for_lattice_thermal_conductivity_at_low_temperatures} and 
consider a momentum-relaxing scattering against defects and phonons through the mean free path $l_{mr}$, as well as electron-electron collisions~\cite{corbino_disk_viscometer_for_2D_quantum_electron_fluids,transverse_magnetosonic_waves_and_viscoelastic_resonance_in_a_two_dimensional_highly_viscous_electron_fluid,bulk_and_shear_viscosities_of_the_two_dimensional_electron_liquid_in_a_doped_graphene_sheet,shear_viscosity_in_interacting_two_dimensional_fermi_liquids} via the collision operator $\Gamma_{ee}$. 
%This is nothing but the Callaway's ansatz~\cite{model_for_lattice_thermal_conductivity_at_low_temperatures}. 
Once we solve Eq.~\eqref{BTE} by the finite element method with the proper boundary conditions~\cite{the_finite_element_method_for_elliptic_problems,alternative_routes_to_electron_hydrodynamics,off_centre_steiner_points_for_delauney_refinement_on_curved_surfaces,supp}, we can compute the drift velocity as ${\bm u}({\bm r}) = (1/\pi) \, \int_{0}^{2\pi} g(\bm r , \theta ) \, (\cos \theta , \sin \theta ) {\, \rm d } \theta $ to get the electric current and the resistance $R$ of a device. We simulate channels of average width $d$, and we express $R$ in units of $R_0 = {\hbar k_F L}/{ne^2 d^2}$, where $L \gg d$ is the length of the channel, long enough to ignore the contacts.

%\textit{Classical dynamics.}

In analogy with conventional fluids, let us first consider electron classical dynamics, where two electrons can collide regardless of their orientation, as depicted in Fig.~\ref{fig:hydrodynamic}(a). Hence, the collision operator reads 
\begin{equation}
\Gamma_{ee} [g] = - \frac{g-g_{ee}}{l_{ee}}\, ,
\label{hydrodynamicDynamics}
\end{equation}
where $l_{ee}$ is the electron-electron mean free path and $g_{ee} = u_x \cos \theta + u_y \sin \theta$ is a polar distribution that moves with the average velocity of the electrons. It is important to note that Callaway's ansatz for the collision operator~\cite{model_for_lattice_thermal_conductivity_at_low_temperatures}, often used in this context, assumes the relaxation of all modes with $m\geq 2$ in the expansion $g = \sum_m \left(c_m \cos m \theta + s_m \sin m \theta\right)$.

Figure~\ref{fig:hydrodynamic}(b) shows the simulated resistance of a channel of width $d$. Note that it is not until $l_{ee} \lesssim d$,  when transport is collective and parallel electrons no longer dominate, that the Gurzhi effect would occur. In such a fluid regime, collisions between electrons keep them from colliding with the edges. Although in the absence of electron-electron collisions, electrons travel a distance $d$, in their presence, they move in shorter steps of length $l_{ee}$. Thus, on average, they propagate a distance $d^2/l_{ee}$ before reaching the edge, reducing the resistance~\cite{minimum_of_resistance_in_impurity_free_conductors}.
Figure~\ref{fig:hydrodynamic}(c) shows the resistance for uniform and crenelated channels with increased corrugation, reproducing the behavior expected for conventional fluids. On the contrary, experiments on electron fluids show no initial increase of the resistance at low temperatures, giving rise to the superballistic paradox. 
In addition, we can also prove that the initial resistance increase would arise, or even worsen, for other edge scattering mechanisms~\cite{graphene_a_nearly_perfect_fluid,boundary_conditions_of_viscous_electron_flow} and $l_{mr}/d$ ratios~\cite{supp}.
As we will discuss, the superballistic paradox originates in the properties of electron-electron collisions.

\begin{figure*}[ht]
    \centering
    \includegraphics[width=1\linewidth]{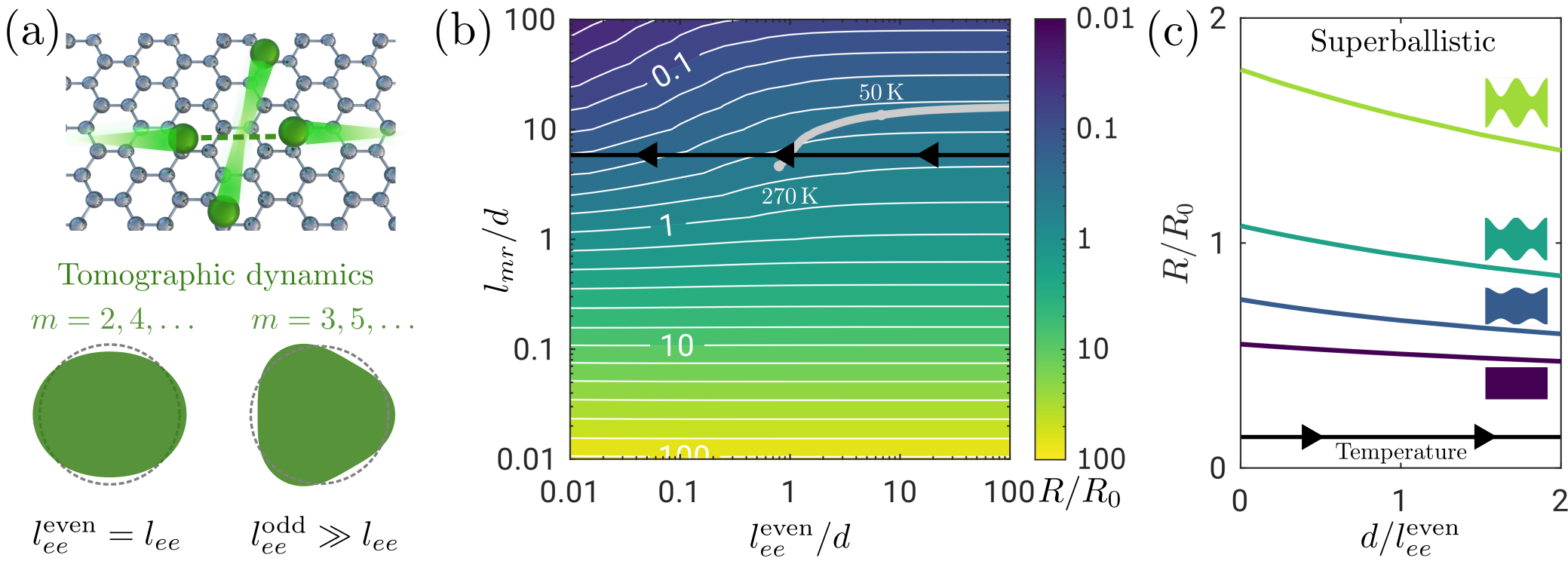}
    \caption{Tomographic dynamics. (a) Only head-on collisions are allowed between electrons, so odd parity modes do not relax. (b) Resistance as a function of the collision rates assuming $l_{ee}^{\rm even} \ll l_{ee}^{\rm odd}$. The gray line shows typical values for a graphene channel of width $d = 200 \, \rm nm$ at $n=0.5\times 10^{12} \, \rm cm^{-2}$~\cite{superballistic_flow_of_viscous_electron_fluid_through_graphene_constrictions,negative_local_resistance_caused_by_viscous_electron_backflow_in_graphene}. (c)     The resistance decreases with electron-electron collisions in uniform and crenelated channels, for $l_{mr} / d = 5$, even for close-to-zero temperatures in agreement with the experiments. 
    }
    \label{fig:tomographic}
\end{figure*}

We now discuss the experimental evidence demonstrating that a lack of measurements at low enough temperatures cannot be responsible for the observed paradox. % Experiments in graphene have studied devices. 
For a typical size $d\sim 0.2 \, \rm \mu m$ and a density of carriers $n = 0.5 \times 10^{12}\, \rm cm^{-2}$, the temperature at which the original Gruzhi condition for superballistic conduction fulfills ($l_{ee} < d $) is $200 \, \rm K$ in graphene devices~\cite{superballistic_flow_of_viscous_electron_fluid_through_graphene_constrictions,superballistic_conduction_in_hydrodynamic_antidot_graphene_superlattices}. Similarly in gallium arsenide heterostructures at $n = 0.25 \times 10^{12} \, \rm cm^{-2}$, 
it is not until $10 \, \rm K$ that we reach $l_{ee} \sim 1 \, \rm \mu m$, the characteristic size of the device~\cite{geometric_control_of_universal_hydodynamic_flow_in_a_two_dimensional_electron_fluid}.
However, in both 2D electron systems, experiments have observed a decrease in the resistance of the superballistic effect for temperatures way below the Gurzhi threshold estimated above. In addition, no increase in the resistance has been observed at closed-to-zero temperatures where measurements were performed. 
This behavior has been confirmed in several experimental set-ups~\cite{boundary_mediated_electron_electron_interactions_in_quantum_point_contacts,temperature_dependence_of_electron_viscosity_in_superballistic_gaas_point_contact,superballistic_electron_flow_through_a_point_contact_in_a_gaalas_heterostructure,superballistic_flow_of_viscous_electron_fluid_through_graphene_constrictions,superballistic_conduction_in_hydrodynamic_antidot_graphene_superlattices,boundary_mediated_electron_electron_interactions_in_quantum_point_contacts,effects_of_electron_electron_scattering_in_wide_ballistic_microcontacts,long_distance_electron_electron_scattering_detected_with_point_contacts,superballistic_electron_flow_through_a_point_contact_in_a_gaalas_heterostructure,quantitative_measurement_of_viscosity_in_two_dimensional_electron_fluids}. Hence, the observed paradox is not exclusive to a particular material or geometry that prompts us to sake a universal explanation.

%\textit{Tomographic dynamics.}

Now, let us get a deeper insight into the differences between conventional and electron fluids to settle the superballistic paradox. Indeed, unlike molecules in conventional fluids that follow a Maxwell distribution, electrons are fermions governed by the Fermi-Dirac statistics. The conservation of energy and momentum in a 2D system and the need for occupied initial and unoccupied final states in scattering events highly restrict collisions. At low temperatures, these are predominantly head-on collisions among electrons facing each other, resulting in the tomographic dynamics~\cite{tomographic_dynamics_and_scale_dependent_viscosity_in_2D_electron_systems,odd_parity_effect_and_scale_dependent_viscosity_in_atomic_quantum_gases,electron_electron_collisions_and_a_new_hydrodynamic_effect_in_two_dimensional_electron_gas,odd_parity_effect_and_scale_dependent_viscosity_in_atomic_quantum_gases,nonequilibrium_relexation_and_odd_even_effect_in_finite_temperature_electron_gases,collective_modes_in_interacting_two_dimensional_tomographic_fermi_liquids,anomalously_long_lifetimes_in_two_dimensional_fermi_liquids,superscreening_by_a_retroreflected_hole_backflow_in_tomographic_electron_fluids,viscosity_of_two_dimensional_electrons,the_hierarchy_of_excitation_lifetimes_in_two_dimensional_fermi_gases,two_dimensional_electron_gases_as_non_newtonian_fluids} depicted in Fig.~\ref{fig:tomographic}(a). To restrict our model to tomographic dynamics, we must split the collision operator as follows
\begin{equation}
\Gamma_{ee} [g] = - \frac{g^{\rm even}-g_{ee}^{\rm even}}{l_{ee}^{\rm even}} - \frac{g^{\rm odd}-g_{ee}^{\rm odd}}{l_{ee}^{\rm odd}}\, ,
\label{tomographicDynamics}
\end{equation}
where $l_{ee}^{\rm even}$ and $l_{ee}^{\rm odd}$ are the mean free path of the even and odd parity modes of the polar distribution $g({\bm r}, \theta)$, respectively. 
Indeed, at low temperatures, the decay length is $l_{ee}^{(m)}  \sim m^4 (T_F / T)^2 l_{ee}^{\rm even}$ for the odd-parity modes $m=3,5\dots$ in the expansion of $g$~\cite{supp,linear_in_temperature_conductance_in_electron_hydrodynamics}. Therefore, at the low-temperature limit, it is enough to consider pure tomographic dynamics where the odd-parity modes obey $l_{ee}^{\rm odd} \gg l_{ee}^{\rm even}$ to explain the paradox resistance behavior~\cite{supp}. 
Notice that head-on collisions do not relax the odd modes, and it is not trivial to determine their impact on the macroscopic properties. Hydrodynamic models akin to the Navier-Stokes equation are blind to the distinction between even and odd parity modes~\cite{supp,alternative_routes_to_electron_hydrodynamics}, and thus, unable to account for the tomographic dynamics. However, the Boltzmann equation distinguishes them, resulting in a crucial theoretical tool to explain the paradox.

The resistance map in Fig.~\ref{fig:tomographic}(b) shows key differences from its classical counterpart of Fig.~\ref{fig:hydrodynamic}(b). Figure~\ref{fig:tomographic}(c) confirms that tomographic dynamics predict no increase in the resistance but a decrease starting at close-to-zero temperatures, in perfect agreement with the experiments, thus solving the paradox. 
Since the electric current is mainly associated with the odd-parity modes, it is no wonder that tomographic dynamics do not exhibit the initial increase in the resistance characteristic of classical dynamics. Indeed, head-on collisions cannot bring the electrons out of their trajectories parallel to the channel~\cite{ballistic_flow_of_two_dimensional_interacting_electrons} and increase the resistance.% However, they reduce the edge scattering~\cite{ballistic_flow_of_two_dimensional_interacting_electrons} and then the resistance even at close-to-zero temperatures.

Theoretical studies predict how tomographic dynamics correct current injectors~\cite{electron_electron_momentum_relaxation_in_a_two_dimensional_electron_gas}, thermoelectric properties~\cite{nonlinear_thermoelectric_probes_of_anomalous_electron_lifetimes_in_topological_fermi_liquids} or the Poiseuille flow~\cite{tomographic_dynamics_and_scale_dependent_viscosity_in_2D_electron_systems}. However, experiments have explained Poiseuille's flow without noticing the subtle difference yet~\cite{visualizing_poiseuille_flow_of_hydrodynamic_electrons}. The search for experimental evidence of tomographic dynamics includes the analysis of magnetotransport in ultrapure devices~\cite{testing_the_tomographic_fermi_liquid_hypothesis_with_high_order_cyclotron_resonance,magnetic_field_suppression_of_tomographic_electron_transport} and of the scaling $ {\rm d} \ln R / {\rm d} \ln T$~\cite{linear_in_temperature_conductance_in_electron_hydrodynamics}. The latter changes with the transport regime~\cite{quantitative_measurement_of_viscosity_in_two_dimensional_electron_fluids} and depends on a previous assumption of the mean free patch temperature-scaling, $l_{ee}\sim T^{-2} $~\cite{viscous_electron_fluids} or $l_{ee} \sim T^{-2} \ln (T_F / T) $~\cite{superballistic_flow_of_viscous_electron_fluid_through_graphene_constrictions}, as well as for the odd parity modes~\cite{superballistic_flow_of_viscous_electron_fluid_through_graphene_constrictions}. Our proposal to solve the superballistic paradox results in clear experimental evidence of the tomographic dynamics that only assume that temperature favors collisions between electrons at low temperatures, a trivial consequence of the Pauli blockade~\cite{viscous_electron_fluids}. 
Furthermore, our approach is not based on a change in the peculiarities of the scaling behavior of $R(T)$. In contrast with previous studies~\cite{linear_in_temperature_conductance_in_electron_hydrodynamics}, under a proper description of edge scattering beyond the non-physical no-slip boundary condition~\cite{supp,limits_of_the_hydrodynamic_no_slip_boundary_condition,boundary_conditions_of_viscous_electron_flow,pressure_driven_diffusive_gas_flows_in_micro_channels_from_the_knudsen_to_the_continuum_regimes}, we demonstrate that classical dynamics lead to an initial increase in the resistance, in agreement with conventional fluids~\cite{knudsen_minimum_original,predicting_the_knudsen_paradox_in_long_capillaries_by_decomposing_the_flow_into_ballistic_and_collision_parts,predicting_the_knudsen_paradox_in_long_capillaries_by_decomposing_the_flow_into_ballistic_and_collision_parts}.

%under the appropriate choice of the boundary conditions~\cite{limits_of_the_hydrodynamic_no_slip_boundary_condition,boundary_conditions_of_viscous_electron_flow,pressure_driven_diffusive_gas_flows_in_micro_channels_from_the_knudsen_to_the_continuum_regimes},} 

\begin{figure}[t]
    \centering
    \includegraphics[width=1\linewidth]{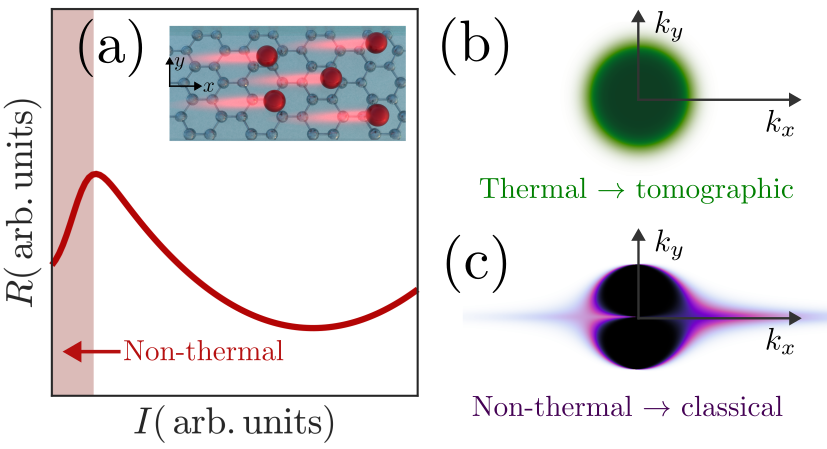}
    \caption{Molenkamp effect scheme. (a) Resistance as a function of the current. (b) The Fermi electron distribution. (c) The non-thermal electron distribution has many electrons traveling parallel to the channel. The collisions are now closer to classical dynamics, resulting in an increasing resistance.     }
    \label{fig:Qualitative2}
\end{figure}

We have already demonstrated that classical and tomographic dynamics yield a Gurzhi effect in 2D electronic systems. However, in the first case, the superballistic effect arises at intermediate temperatures. In contrast, in the second scenario, it already starts at close-to-zero temperatures. Since electron dynamics is tomographic at low temperatures, we have confirmed that the reduction of the resistance starts at zero temperature, in agreement with experiments~\cite{boundary_mediated_electron_electron_interactions_in_quantum_point_contacts,temperature_dependence_of_electron_viscosity_in_superballistic_gaas_point_contact,superballistic_electron_flow_through_a_point_contact_in_a_gaalas_heterostructure,superballistic_flow_of_viscous_electron_fluid_through_graphene_constrictions,superballistic_conduction_in_hydrodynamic_antidot_graphene_superlattices,boundary_mediated_electron_electron_interactions_in_quantum_point_contacts,effects_of_electron_electron_scattering_in_wide_ballistic_microcontacts,long_distance_electron_electron_scattering_detected_with_point_contacts,superballistic_electron_flow_through_a_point_contact_in_a_gaalas_heterostructure,quantitative_measurement_of_viscosity_in_two_dimensional_electron_fluids}.  
However, a different behavior was found in Molenkamp's experiment~\cite{electronic_poiseuille_flow_in_hexagonal_boron_nitride_encapsulated_graphene_FETS,hyrdodynamic_electron_flow_in_high_mobility_wires,electron_electron_scattering_induced_size_effects_in_a_two_dimensional_wire,negative_differential_resistance_of_viscous_electron_flow_in_graphene}, where, instead of increasing temperature, high electric currents are used. In such a case, the rate of electron-electron collisions rises due to the applied current. We schematically show Molenkamp's experimental results in Fig.~\ref{fig:Qualitative2}(a). Remarkably, and contrary to the measurements obtained for increasing temperatures, Molenkamp's work reported an initial increase in the resistance at low currents, namely, at low electron-electron collision rates. After this experiment, Gurzhi adapted their original predictions for the superballistic effect in a metal ($l_{ee} < d$), for a 2D tomographic electron system to demonstrate a decrease in the resistance for an easier condition to fulfill, namely $l_{ee} \sqrt{T/T_F} < d$~\cite{electron_electron_collisions_and_a_new_hydrodynamic_effect_in_two_dimensional_electron_gas}. The latter would lead to a decrease in the resistance for lower temperatures consistently with Molenkamp's experiment. Once again, for graphene at $n = 0.5\times 10^{12} \, \rm cm^{-2}$ the new condition $l_{ee} \sqrt{T / T_F} \sim d\sim 0.2 \, \rm \mu m$ would be valid above $70 \, \rm K$~\cite{superballistic_flow_of_viscous_electron_fluid_through_graphene_constrictions,negative_local_resistance_caused_by_viscous_electron_backflow_in_graphene}. However experiments do not reveal any data compatible with a Knudsen minimum at such temperature neither. Remarkably, our simulations under the tomographic approach demonstrate that the resistance decreases regardless of the condition $l_{ee} \sqrt{T/T_F} < d$, which was not possible to spot by analytic calculations~\cite{electron_electron_collisions_and_a_new_hydrodynamic_effect_in_two_dimensional_electron_gas}.

The peculiarity of the Molenkamp effect, with a starting increase in the resistance, lies in the lack of thermal equilibrium. When the device's temperature rises, electrons still follow the Fermi distribution. However, if a high current is applied, it mainly accelerates those electrons traveling parallel to the channel, and its distribution is no longer thermal. Tomographic dynamics is derived under the assumption of 2D electrons within the Fermi distribution, as shown in Fig.~\ref{fig:Qualitative2}(b). Thus, it is no wonder its failure to describe Molenkamp's results where the distribution is non-thermal, like the one depicted in Fig.~\ref{fig:Qualitative2}(c). Notice that an arbitrary electron will mostly collide with another electron moving parallel to the channel since the distribution is much broader in that direction~\cite{supp}. Indeed, these new collisions overshadow the head-on tomographic ones for sharper distribution functions. To support our analysis we solve the collision integrals~\cite{anomalously_long_lifetimes_in_two_dimensional_fermi_liquids,the_hierarchy_of_excitation_lifetimes_in_two_dimensional_fermi_gases} for the non-thermal distribution and demonstrate that $l_{ee}^{\rm even} $ and $l_{ee}^{\rm odd}$ are comparable in that case, so that tomographic dynamics ($l_{ee}^{\rm odd} \gg l_{ee}^{\rm even}$) no longer applies~\cite{supp}. Instead, the obtained behavior is close to the one described by classical dynamics that gives rise to an increasing resistance for low frequent electron-electron collisions, as previously demonstrated.
The application of even higher currents and the activation of more electronic collisions eventually lead to the thermalization of the Fermi distribution. The latter and the increase in $d/l_{ee}^{\rm even}$ contribute to the decreasing resistance observed in Molenkamp's experiment. The Molenkamp effect, which seemed to be an exception in our theory, reinforces our analysis. Physical scenarios without thermal equilibrium opened a window for classical dynamics.

%\textit{Conclusions.} 

In conclusion, we demonstrate that the petty details of electron dynamics dramatically alter its electrical properties. At low temperatures, classical dynamics lead to an increase of resistance upon rising temperature, while tomographic dynamics lead to a decreasing resistance. Experiments show a decrease in the resistance even at close-to-zero temperatures, with no Knudsen minimum, which is in total agreement with our tomographic simulations. Therefore, this electron's peculiarity signals the occurrence of tomographic dynamics, strengthening the superballistic effect and enabling its application to low-dissipation devices. Our theory explains the different behavior between electrons and classical fluids and, by considering current-driven non-thermal phenomena, the Molenkamp effect. 
We ultimately solve the superballistic paradox, which is a consequence of electrons not being classical particles but fermions.

\begin{acknowledgments}

%\textit{Acknowledgments.} 

We thank R. Brito for discussions. This work was supported by the "(MAD2D-CM)-UCM" project funded by Comunidad de Madrid, by the Recovery, Transformation, and Resilience Plan, and by NextGenerationEU from the European Union and Agencia Estatal de Investigaci\' {o}n of Spain (Grant PID2022-136285NB-C31). J.~E.-A. acknowledges support from the Spanish Ministerio de Ciencia, Innovaci\' {o}n y Universidades (Grant FPU22/01039).

\end{acknowledgments}

\bibliography{biblio} 

\newpage 

%%%%%%%%%% Merge with supplemental materials %%%%%%%%%%
\pagebreak
\widetext
\begin{center}
\textbf{\large Supplemental Materials\\ Superballistic paradox in electron fluids:
Evidence of tomographic transport}
\end{center}
%%%%%%%%%% Merge with supplemental materials %%%%%%%%%%
%%%%%%%%%% Prefix a "S" to all equations, figures, tables and reset the counter %%%%%%%%%%
\setcounter{equation}{0}
\setcounter{figure}{0}
\setcounter{table}{0}
\setcounter{page}{1}
\makeatletter
\renewcommand{\theequation}{S\arabic{equation}}
\renewcommand{\thefigure}{S\arabic{figure}}
%\renewcommand{\bibnumfmt}[1]{[S#1]}
%\renewcommand{\citenumfont}[1]{S#1}
%%%%%%%%%% Prefix a "S" to all equations, figures, tables and reset the counter %%%%%%%%%%

\section{Boltzmann transport equation \label{SI_BTE}}

We solve the Boltzmann transport equation to describe ballistic, tomographic, and hydrodynamic transport. Let us follow a similar procedure to Ref.~\cite{alternative_routes_to_electron_hydrodynamics} to understand its meaning. We characterize the electrons as semiclassical particles moving in two dimensions with a well-defined position $\bm r = (x, y)$ and wave vector $\bm k = (k_x , k_y)$. 
%\textcolor{red}{This semiclassical description is valid, provided that there are many transmission channels and quantization effects are negligible, as it happens in devices where the Gurzhi effect is studied}. 
We consider an isotropic band structure where the electron's velocity is $\bm v = \hbar \bm k / m $, and $m$ is the cyclotron effective mass in a parabolic dispersion relation. We can also describe a linear band structure with constant velocity $\bm v = v_F \bm k / k $, like, for example, that of single-layer graphene with the Fermi velocity  $v_F \simeq 10^6 \rm \, m/ s$. In this case, $m$ is the cyclotron's effective mass. We will consider the electron distribution $f ( \bm r , \bm k )$, such that $ f (\bm r , \bm k ) \mathcal{N} / 4\pi^2$ gives how many electrons at position $\bm r$ have a wave vector $\bm k$, being $\mathcal{N} = 4$ the number of equivalent valleys and spins in graphene. In the steady state, the electrons in a 2D system obey the following Boltzmann transport equation
\begin{equation}
\bm v \cdot \nabla_{\bm r} f + \frac{e}{\hbar} \nabla V  \cdot \nabla_{\bm k} f = \Gamma_{mr} [ f ]  + \Gamma_{ee} [ f ] \ , 
\label{fBTE}
\end{equation}
where we consider that electrons are subject to an electric potential $V (\bm r)$. We also consider a collision operator to describe electron-phonon and electron-defect scattering ($\Gamma_{mr}$) and electron-electron scattering ($\Gamma_{ee}$). In the relaxation time approximation, electron scattering with defects and phonons yields the following collision operator
\begin{equation}
\Gamma_{mr}[f] = -\frac{f-f_e}{\tau_{mr}}\, , 
\end{equation}
with a constant relaxation time $\tau_{mr}$. They eventually drive the electrons to the Fermi equilibrium distribution $f_e (\bm r , \bm k)  = 1 $ if $k < k_F$ and is $0 $ otherwise. Here, the Fermi wave number $k_F= \sqrt{4\pi n/\mathcal{N}}$ is determined by the carrier density, assumed to be constant in experiments of electron hydrodynamics. There, it is set using a back gate, and we assume $n = (\mathcal{N} / 4\pi^2) \int_{\bm k} f(\bm r , \bm k) \, {\rm d} \bm k  $. Electrons also undergo collisions with other electrons
\begin{equation}
\Gamma_{ee}[f] = -\frac{f-f_{ee}^{\rm even}}{\tau_{ee}^{\rm even}} - \frac{f-f_{ee}^{\rm odd}}{\tau_{ee}^{\rm odd}}\, , 
\end{equation}
in processes that conserve the total electron momenta. These collisions do not take electrons to the Fermi distribution, but rather to a shifted distribution $f_{ee}(\bm r, \bm k )  = f_e (\bm r, \bm k - \bar{\bm k} )$ that has the same momenta $\bar{\bm k}(\bm r)  = \int_{\bm k} \bm k \, f(\bm r, \bm k )  \, {\rm d} \bm k / \int_{\bm k} f(\bm r, \bm k )  \, {\rm d} \bm k  $ as the original one. We consider two relaxation times for the even and odd parity modes in the polar expansion of $f$, enabling us to describe classical and tomographic dynamics.
The validity of the Boltzmann equation to describe electron dynamics has been previously discussed~\cite{alternative_routes_to_electron_hydrodynamics}, and it holds, provided that the applied field is small enough, which is a sensible consideration in experiments. It also assumes that there are many transmission channels, so quantization effects are negligible for experiments on superballistic conduction~\cite{alternative_routes_to_electron_hydrodynamics}. 
We can simplify the problem by writing $\bm k = (k \cos \theta , k \sin \theta )$ in polar coordinates and defining the following polar distributions
\begin{align}
g (\bm r , \theta ) = &\frac{v_F}{k_F} \int_{0}^\infty \left[ f(\bm r , k , \theta) - f_e(\bm r , k , \theta) \right] \, {\rm d} k \ , \\
 g_{ee} (\bm r , \theta ) = & \frac{v_F}{k_F} \int_{0}^\infty \left[ f_{ee}(\bm r , k , \theta) - f_e(\bm r , k , \theta) \right] \, {\rm d} k   \, .
\end{align}

We also define the drift velocity $\bm u = (u_x, u_y) $ as follows 
\begin{equation}
\bm u (\bm r ) = \frac{\int \bm v \, f(\bm r , \bm k ) \, {\rm d} \bm k }{\int  f(\bm r , \bm k ) \, {\rm d} \bm k}\simeq \frac{v_F}{k_F \int  f(\bm r , \bm k ) \, {\rm d} \bm k } \int \left(\begin{matrix}
\cos \theta \\ \sin \theta
\end{matrix} \right)  \, f(\bm r , \bm k ) \, {\rm d} \bm k  \simeq \frac{1}{\pi} \int_{0}^{2\pi}\left(\begin{matrix}
\cos \theta \\ \sin \theta
\end{matrix} \right)  \,  g (\bm r, \theta ) \, {\rm d} \theta \, ,
\end{equation}
Moreover, we assume that transport phenomena occur near the Fermi surface $| g (\bm r, \theta )  | \ll v_F$, meaning that the term inside the integral is non-zero only near $k_F$, a reasonable hypothesis for experiments even at the high currents $\sim 10 \, \rm \mu A$ used in the region of the Molenkamp effect that we study. After some algebra, we find 
\begin{equation}
g_{ee}(\bm r, \theta )  \simeq u_x(\bm r ) \cos \theta + u_y(\bm r) \sin \theta \, . 
\end{equation}
We can now integrate Eq.~\eqref{fBTE} from $k=0$ to $\infty$ to derive 
\begin{equation}
\left(\begin{matrix}
\cos \theta \\ \sin \theta
\end{matrix} \right) 
\cdot \nabla_{{\bm r}} \left( g- \frac{e V}{ \hbar k_F} \right)  = -\frac{g}{l_{mr}} - \frac{g^{\rm even}-g_{ee}^{\rm even}}{l_{ee}^{\rm even}} - \frac{g^{\rm odd}-g_{ee}^{\rm odd}}{l_{ee}^{\rm odd}} \, ,
\end{equation}  
where no hypothesis on the ratios between $l_{mr} = v_F \tau_{mr}$, $l_{ee}^{\rm even} = v_F \tau_{ee}^{\rm even}$, $l_{ee}^{\rm odd} = v_F \tau_{ee}^{\rm odd}$ and the size $d$ of the device has been made. This is the main equation to be used from now on. 

\section{Numerical methods \label{SI_numerical}}

We solve the Boltzmann equation with a conformal Galerkin finite element method~\cite{the_finite_element_method_for_elliptic_problems,alternative_routes_to_electron_hydrodynamics}. We write the solution in a uniform channel as 

\begin{equation}
    g(\bm r , \theta ) = \sum_{n=1}^N \sum_{m=1}^M g_{nm} \phi_n ({\bm r}) \varphi_m(\theta )\ .
    \label{gExpansion1D}
\end{equation}

We consider $\lbrace \phi_n \rbrace_{n=1}^N$ as a basis of tent functions defined on the $[-d/2,d/2]$ interval, and the products of adjacent tent functions. We also take the angular elements $\lbrace \varphi_m \rbrace_{m=1}^M$, which are the tent functions defined on the periodic $[0, 2\pi ) $ interval. We achieve convergence with $N = 41$ and $M = 32$ or $M = 128$ for the more exigent no-slip condition. In a uniform channel, the Boltzmann equation reduces to
\begin{equation}
\sin \theta \,\frac{\partial g }{\partial y} - \frac{e}{\hbar k_F}\, \frac{\partial V }{\partial x} \cos \theta 
 = -\frac{g}{l_{mr}} - \frac{g^{\rm even}-g_{ee}^{\rm even}}{l_{ee}^{\rm even}} - \frac{g^{\rm odd}-g_{ee}^{\rm odd}}{l_{ee}^{\rm odd}} \, .
\end{equation}  

We write the weak formulation of this equation and substitute the expansion for $g$, with the resulting linear system being solved using an iterative least-square method in Matlab. We also include the boundary condition for scattered electrons at the edges. We generalize the method to other 2D geometries by writing 
\begin{equation}
    g(\bm r , \theta ) = \sum_{n=1}^N \sum_{m=1}^M g_{nm} \phi_n (\bm r) \varphi_m(\theta ) \hspace{1cm} V(\bm r) = \sum_{n=1}^N V_n \phi_n (\bm r) \ .
    \label{gExpansion2D}
\end{equation}

For each crenelated channel geometry, the unit cell has a period $d$ and a mean width $d$, meaning that the edges of the channels are modulated by the functions $y = [1 \pm A  \cos  \left( 2\pi x / d \right)]\,d$, where $A = 0$ gives the uniform channel and $A = 0.1$, $0.2 $ and $0.3$ account for crenelated channels of increasing corrugation. We consider $\lbrace \phi_n \rbrace_{n=1}^N$ to be the set of tent functions and their bubbles on a triangular mesh defined using a Delaunay triangulation~\cite{off_centre_steiner_points_for_delauney_refinement_on_curved_surfaces} with sizes $h < 0.15 \, d$. We also consider the angular elements $\lbrace \varphi_m \rbrace_{m=1}^M$, which are the tent functions defined on the periodic $[0, 2\pi ) $ interval. We take $M = 16$ for convergence. We set a constant density of carriers and use periodic boundary conditions at the limits of the simulated cell $g$ distribution at the limits of the simulated cell for the periodic channel. In both cases, once we determine $g (\bm r, \theta ) $ and the potential $V (\bm r)$, we calculate the velocity field $\bm u (\bm r)$. We also compute the current density and numerically integrate its profile to compute the channel's total current and the resistance $R$. We use another conformal Galerkin finite element method for the crenelated geometry and the supplementary Navier-Stokes model. This time, there is no dependence in $\theta$, and we use the analytic expression for the velocity profile in a channel to set the velocity at the edges of the simulation, away from the studied region. We implement a partial slip boundary condition with the slip length given by Eq.~\eqref{slipLength} at the edges, write the weak formulation of the problem, and solve the subsequent linear system in Matlab.

\section{Edge scattering}

\begin{figure}[ht]
    \centering
\includegraphics[width=0.5\linewidth]{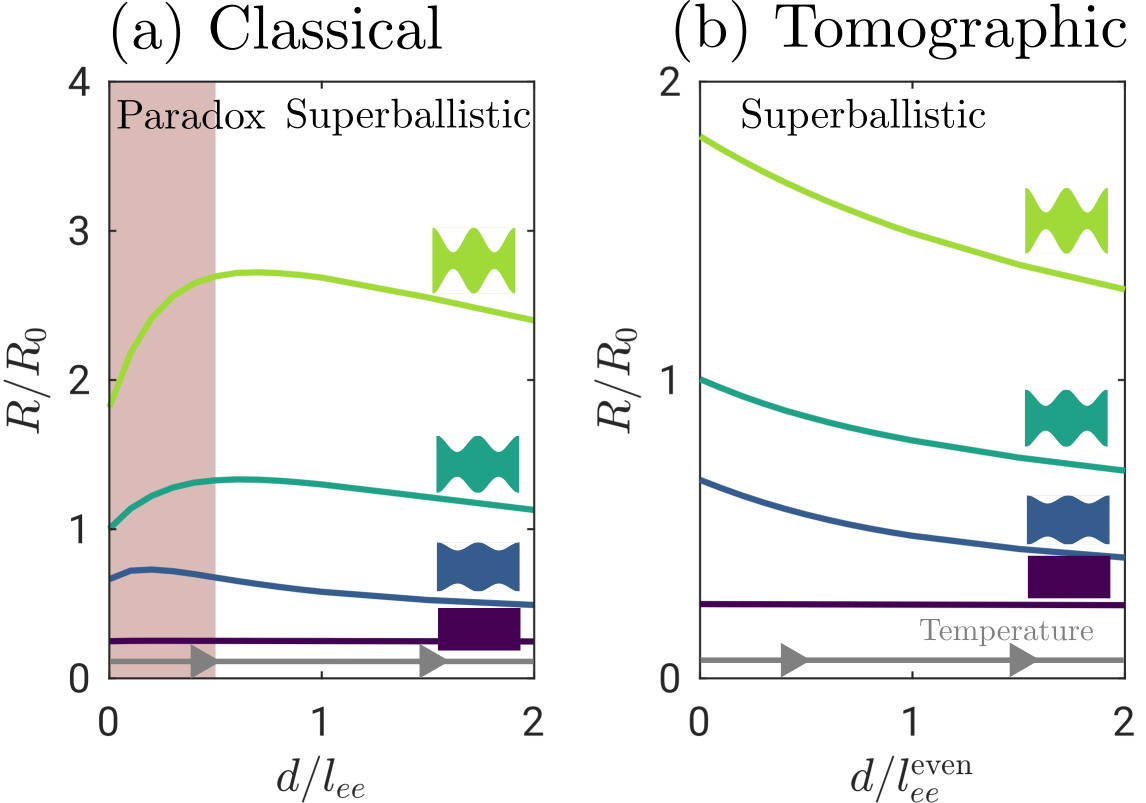}
    \caption{The conclusions on the superballistic paradox are robust regardless of the edge scattering mechanism. (a)-(b) Resistance as a function of $l_{ee}^{-1}\sim T^{2}$ in the same physical scenarios considered in Fig.~\ref{fig:hydrodynamic}(c) and Fig.~\ref{fig:tomographic}(c) respectively, but under specular edge scattering.  }
    \label{fig:edge}
\end{figure}
Electrons are subject to edge scattering at the edges of the device. One edge scattering mechanism is diffusive scattering, where electrons reaching the edge of the device scatter uniformly in all directions, meaning that the polar distribution function evaluated at the edges is
\begin{equation}
g(\theta ) = 0 \hspace{0.5cm} 0 < \theta < \pi 
\end{equation}
for all reflected electrons, namely when $0 < \theta < \pi $ or with the corresponding rotation for a specific edge orientation. Diffusive scattering is often associated with a rough, disordered edge, and it is the condition that we use in the main manuscript. However, it is possible to explore another set of boundary conditions where the polar distribution accounts for partial reflection 
\begin{equation}
g(\theta ) = g(-\theta ) +  \mathcal{D}  \sin \theta  \left[ g(-\theta ) - \frac{2}{\pi}\,\sin\theta\int_{0}^{\pi} \sin^2 \theta^{\prime} g(-\theta^{\prime}) \, {\rm d} \theta^{\prime} \right]\ \hspace{0.5cm} 0 < \theta < \pi  ,
\label{boundary}
\end{equation}
where $\mathcal{D} = \sqrt{\pi} h^2 he k_F^3 \lesssim 1$, $h$ is the mean height of the bumps in the edge, and $h'$ is its correlation length. Figure~\ref{fig:edge} shows the calculations for specular edge scattering $\mathcal{D} = 0$, associated with a smooth edge, where all electrons reflect. The uniform channel generates a uniform velocity field ${\bm \nabla}^2 u = 0$, hiding hydrodynamic effects. However, we can still analyze the other geometries. We observe that classical and tomographic dynamics are still associated with close-to-zero temperature increases and decreases in resistance. Intermediate edge scattering mechanisms lay between the diffusive and specular extreme scenarios, ensuring our conclusions are robust regardless of the edge scattering mechanism. \\

\begin{figure}[ht]
    \centering    \includegraphics[width=0.5\linewidth]{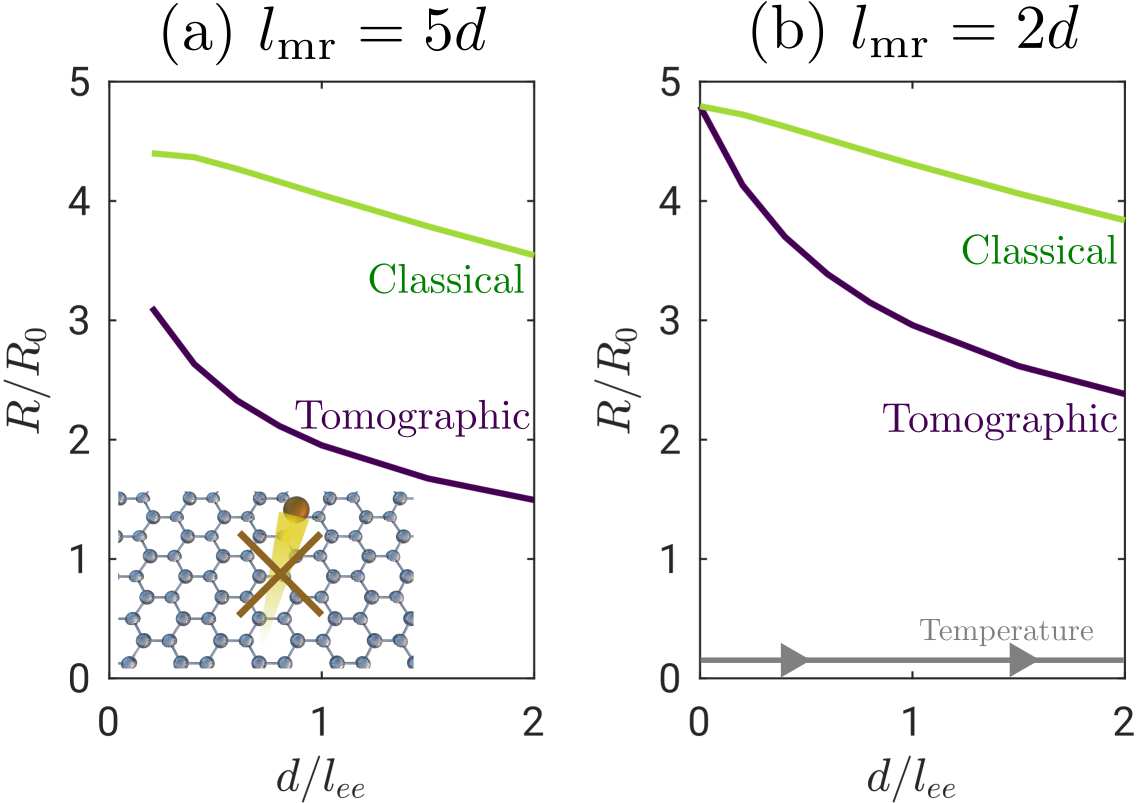}
    \caption{The non-physical no-slip boundary condition would predict different results. (a) Resistance decreases with electron-electron scattering. The inset shows the no-slip boundary conditions, where no incident electrons are allowed at the edge. (b) The same decrease for another scattering rate against impurities and phonons.  }
    \label{fig:noSlip}
\end{figure}

In the deep hydrodynamic regime, where $l_{ee} \ll d $, conventional fluids are usually described by the Navier-Stokes equation with a no-slip boundary condition, $\bm u = 0$. We find the results in Fig.~\ref{fig:noSlip} by solving the Boltzmann equation for a no-slip boundary condition, with $g(\theta ) = 0 $ for $0 \leq \theta < 2\pi $ at the edges. The no-slip boundary condition matches the scaling behavior, in agreement with the scaling laws in~\cite{linear_in_temperature_conductance_in_electron_hydrodynamics}. However, the no-slip condition fails at small $d / l_{ee}$ despite the relevance of the scaling laws therein derived. Indeed, the no-slip boundary condition, which imposes an artificial constraint for incident electrons before their scatter against the edge, cannot be used in the ballistic regime~\cite{limits_of_the_hydrodynamic_no_slip_boundary_condition,boundary_conditions_of_viscous_electron_flow,pressure_driven_diffusive_gas_flows_in_micro_channels_from_the_knudsen_to_the_continuum_regimes}. Therefore, we should use more realistic descriptions of boundary scattering, such as the diffusive or partially specular edges. Doing so, we find a quantitative difference between classical and tomographic dynamics. In contrast to no-slip boundary conditions, our results also explain the Knudsen minimum in conventional fluids~\cite{knudsen_minimum_original} and the Molenkamp effect~\cite{hyrdodynamic_electron_flow_in_high_mobility_wires}.

\section{Collisions against defects}
\begin{figure}[ht]
    \centering    \includegraphics[width=0.5\linewidth]{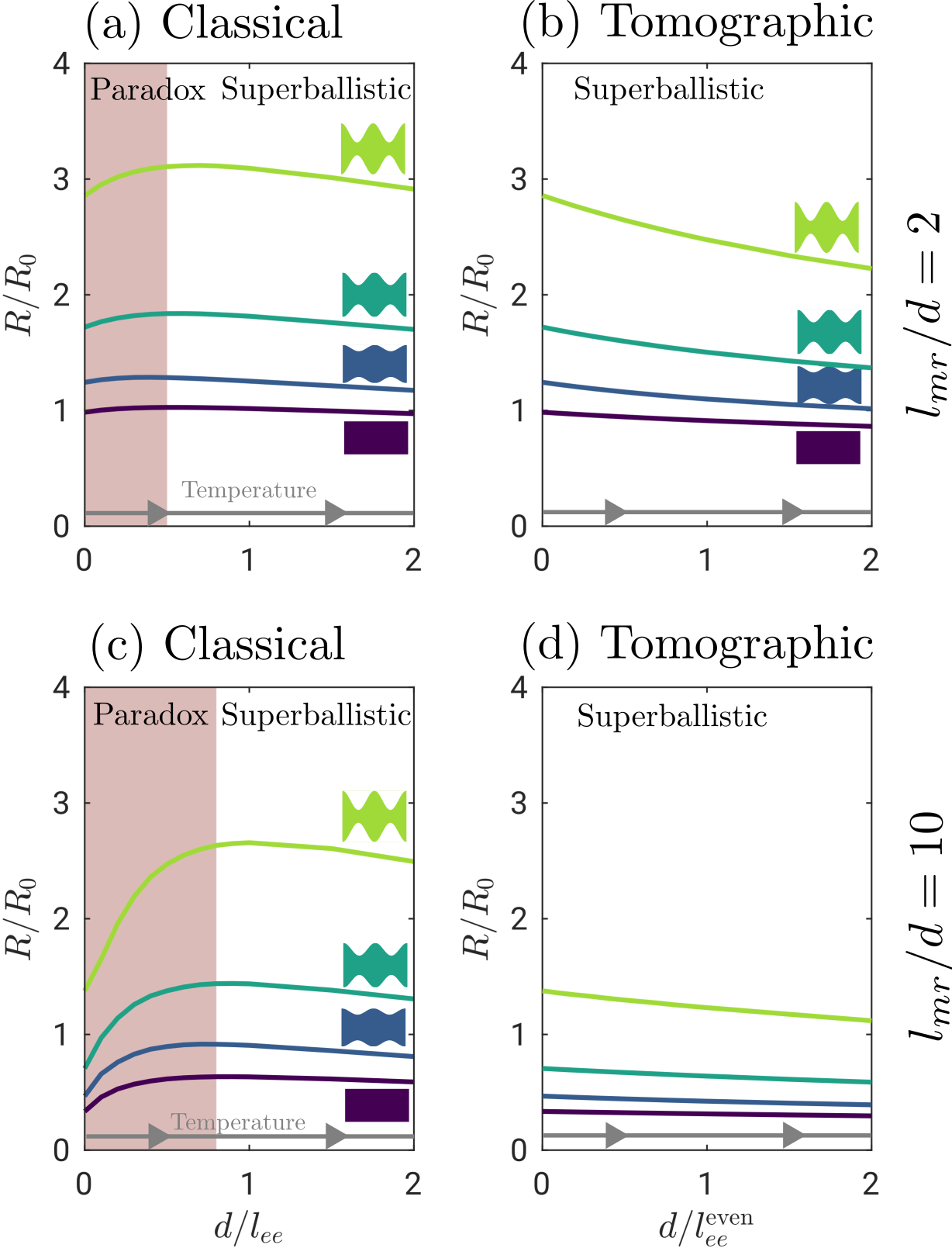}
    \caption{The conclusions on the superballistic paradox are robust regardless of the collisions against defects. (a) The equivalent of classical dynamics in Fig.~\ref{fig:hydrodynamic} and (b) tomographic dynamics in Fig.~\ref{fig:tomographic} for diffusive edge scattering and $l_{mr} / d = 2$. (c) Classical and (d) tomographic dynamics for $l_{mr} / d = 10$.  }
    \label{fig:defects}
\end{figure}

Compared to conventional fluids, electrons suffer collisions against defects and phonons, resulting in a finite $l_{mr} /d $. Figures~\ref{fig:hydrodynamic} and~\ref{fig:tomographic} show that the superballistic paradox arises for non-diffusive $l_{mr} / d \gtrsim 1$, and, particularly at $l_{mr} / d \to \infty $, meaning that scattering against defects is not responsible for the paradox. In the main text, we chose $l_{mr} / d = 5$ for the crenelated channel. Here, we show that it is not critical to our conclusion, with Fig.~\ref{fig:defects} showing analogous results for another value of $l_{mr} / d$.

\section{Between classical and tomographic dynamics}

\begin{figure}[ht]
    \centering
    \includegraphics[width=0.65\linewidth]{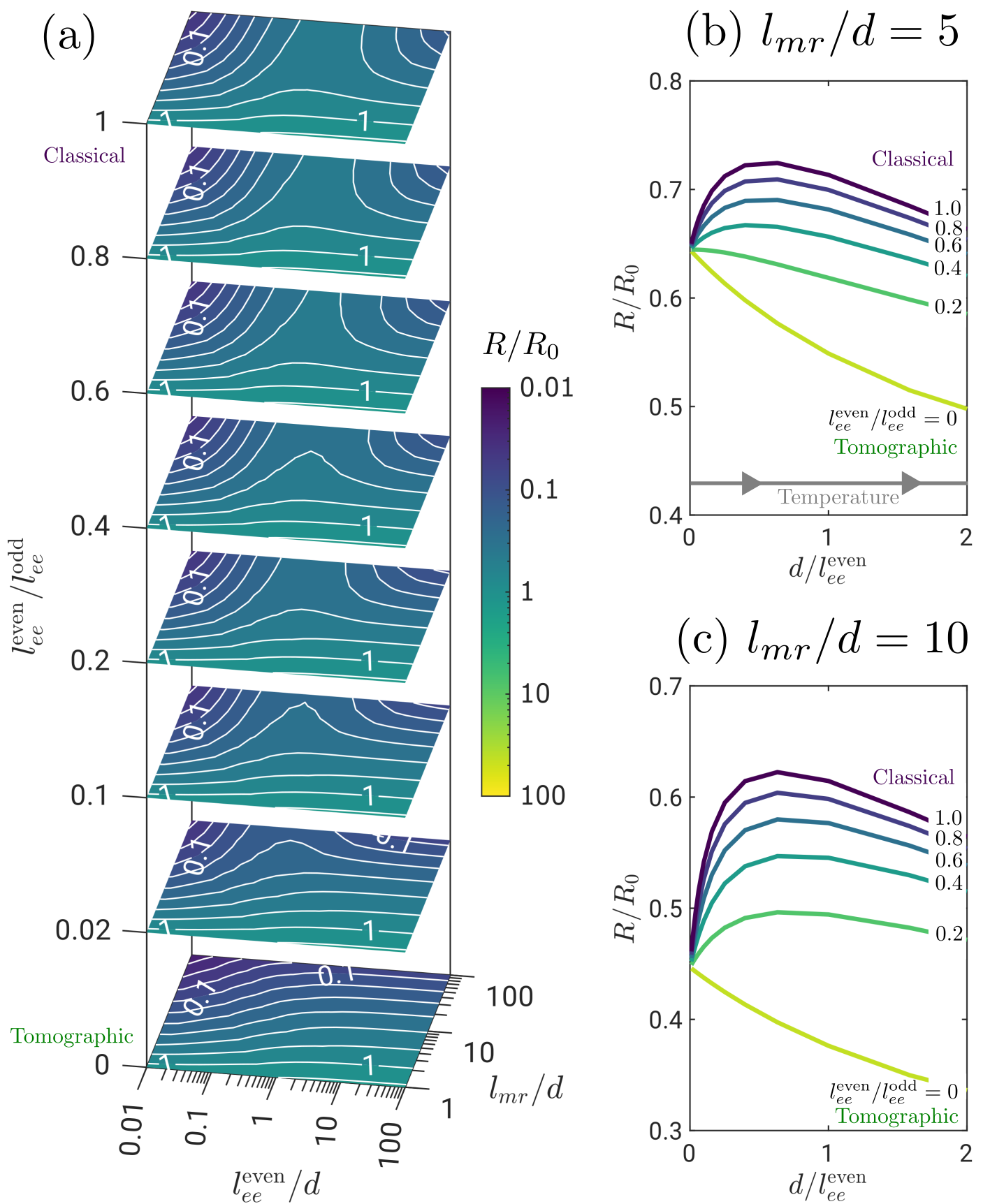}
    \caption{Extended calculations of the resistance as a function of $l_{mr}$, $l_{ee}^{\rm even}$ y $l_{ee}^{\rm odd}$. (a)~Intermediate resistance maps between pure classical and tomographic dynamics, the vertical axis is not linearly spaced. (b)~ resistance as a function of the scattering rate against electrons, transitioning between classical and tomographic dynamics for $l_{mr} / d = 5$ and (c)~$l_{mr}/d = 10$.   \label{fig:Cuts}}    
\end{figure}

Considering the classical $l_{ee}^{\rm even} / l_{ee}^{\rm odd} = 1$ and tomographic $l_{ee}^{\rm even} / l_{ee}^{\rm odd} = 0$ limits is enough to solve the superballistic paradox. This approach explains whether the resistance increases or decreases in the low-temperature limit. In this section, however, we go one step beyond and discuss some calculations for custom combinations of $l_{mr}$, $l_{ee}^{\rm even}$ y $l_{ee}^{\rm odd}$. Figure~\ref{fig:Cuts}(a) shows the intermediate scenarios, transitioning between Fig.~\ref{fig:hydrodynamic}(b) and Fig.~\ref{fig:tomographic}(b). Last, Figs.~\ref{fig:Cuts}(b)--(c) shows the resistance as a function of the scattering rate for several ratios $l_{ee}^{\rm even} / l_{ee}^{\rm odd}$. For $l_{ee}^{\rm even} / l_{ee}^{\rm odd} = 0.4$ there is still a sheer Knudsen minimum, qualitatively resembling the fully classical approach, and even for $l_{ee}^{\rm even} / l_{ee}^{\rm odd} = 0.2$ there is a slight increase and a Gurzhi plateau which is not observed in experiments. The set of ratios $l_{ee}^{\rm even} / l_{ee}^{\rm odd}$ for which the resistance initially decreases narrows as we reduce the collision rates. Therefore, in pure materials with $l_{mr} \gtrsim d$, it is only for $l_{ee}^{\rm even} / l_{ee}^{\rm odd} \ll 1$ that there is a decrease in the resistance, and pretty much any finite $l_{ee}^{\rm even} / l_{ee}^{\rm odd}$  would result in increasing resistance. \\

We can last explore the intermediate regimes using this expression for the change in the resistance
\begin{equation}
R(T) - R(0) \simeq \left[ \frac{{\rm d } R}{{\rm d}  (d/l_{ee} ) } \right]_{\rm classical}\, \Delta \left( \frac{d}{l_{ee}^{\rm odd}} \right) + \left[ \frac{{\rm d } R}{{\rm d}  (d/l_{ee}^{\rm even} ) } \right]_{\rm tomographic}\, \left[ \Delta \left( \frac{d}{l_{ee}^{\rm even}} \right) - \Delta \left( \frac{d}{l_{ee}^{\rm odd}} \right) \right]   \, ,
\end{equation}
which results from fundamental relationships with derivatives. This expression is valid for low temperatures and does not include the phonons' role. Since calculations show that $l_{ee}^{\rm even} / l_{ee}^{\rm odd} \sim (T / T_F)^2$~\cite{tomographic_dynamics_and_scale_dependent_viscosity_in_2D_electron_systems}, one of the terms dominates in the low temperature limit
\begin{equation}
R(T) - R(0) \simeq  \left[ \frac{{\rm d } R}{{\rm d}  (d/l_{ee}^{\rm even} ) } \right]_{\rm tomographic}\, \Delta \left( \frac{d}{l_{ee}^{\rm even}} \right)   \, ,
\end{equation}

Namely, to study the superballistic paradox, we do not need to include the correction due to the finite values of $l_{ee}^{\rm odd}$. The low-temperature behavior, and whether the resistance starts increasing or decreasing, is sufficiently characterized under the deep tomographic approach $\Delta (d/l_{ee}^{\rm even} ) \gg \Delta (d/l_{ee}^{\rm odd}) $. 

\section{Polar distribution functions}

We can further analyze electron transport by looking at the $g(\bm r , \theta)$ polar distribution functions~\cite{magnetic_field_suppression_of_tomographic_electron_transport,collective_modes_in_interacting_two_dimensional_tomographic_fermi_liquids}, which will also be helpful for the discussion of non-thermal phenomena. Let us start by deriving a closed expression for $g(0,\theta)$ at the center of a channel parallel to the $x$ axis with a diffusive edge in the absence of electron-electron collisions. The distribution counts how many electrons, above the Fermi equilibrium distribution, move at an angle $\theta$. If these electrons come directly from the wall, the distance that they have traveled reads 
\begin{equation}
h = \frac{d}{2|\sin \theta|} \, . 
\end{equation}

However, electrons suffer collisions against defects and phonons, with a mean free path $l_{mr}$. The random character of these collisions results in an exponential decay law. Therefore, the average length traveled by electrons is not $h$ but rather 
\begin{equation}
h^* = l_{mr} \left[ 1 - \exp\left( -\frac{d}{2 l_{mr} | \sin \theta |} \right) \right] \, . 
\end{equation}
Collisions avoid the divergence in $h^*$ when $\theta \to 0$. Similarly to Drude's model, the excess electrons are proportional to the average distance $h^*$ that they traveled and the projection of the accelerating field in the direction of $\theta$, given by $\cos \theta$. Therefore 
\begin{equation}
g(0, \theta ) = g_0 \,  \cos \theta \left[ 1 - \exp\left( - \frac{d}{2 l_{mr} | \sin \theta |} \right) \right]\, ,
\label{gAnayltical}
\end{equation}
where $g_0$ is a constant with the units of $g$. In the diffusive regime $l_{mr} \ll d$, the distribution is $g(\theta) = g_0 \cos \theta$, but, as we move into the ballistic regime, there begins to be an accumulation of electrons traveling parallel to the channel. Consequently, $g(\theta ) $ features a sharp peak around $\theta = 0$ approximately given by $g(\theta ) \propto 1/|\tan \theta |$ due to the excess electrons traveling parallel to the geometry.\\

\begin{figure}[ht]
    \centering
    \includegraphics[width=\linewidth]{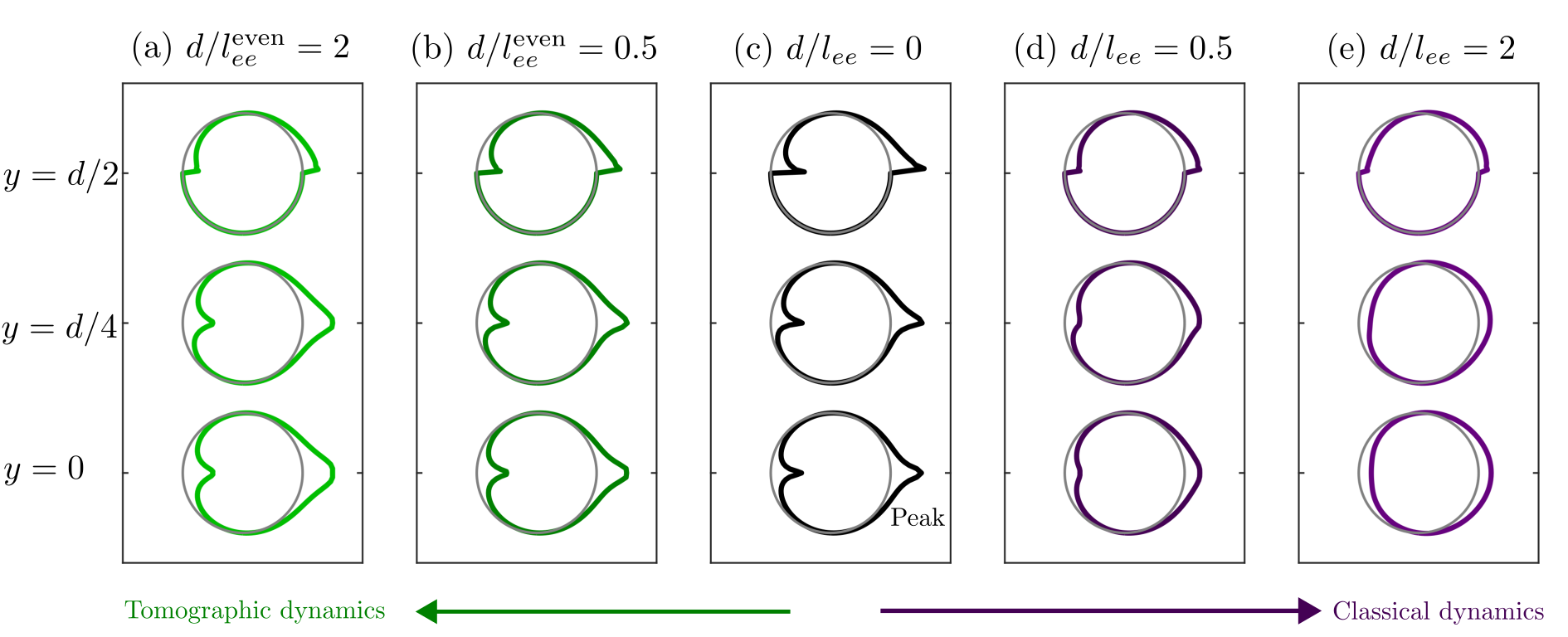}
    \caption{Distribution functions. We depict the polar distribution of $g(y,\theta )$, which gives the excess of electrons traveling at angle $\theta$ and several distances $y$ from the center of the channel, for $l_{mr} / d = 5$. We consider (a)-(b) tomographic dynamics, (c) no collisions between electrons, and (d)-(e) classical dynamics.    }
    \label{fig:Distribution}
\end{figure}

Now, we can use numerical simulations to include the effect of collisions between electrons. Figure~\ref{fig:Distribution} shows the distribution function $g(\theta)$ under tomographic and classical dynamics. In the limiting case of no electron-electron collisions, we can see the peak predicted by Eq.~\eqref{gAnayltical} at the center of the channel, with very similar features at other channel positions. The symmetry of the problem ensures that the $g(\theta)$ function at the center of the channel is only composed of odd-parity modes $\cos(m \theta) $ with odd $m$. Outside the center, the distribution is still similar and mostly consisting of even parity modes. Therefore, it is no wonder that classical dynamics, where the odd-parity modes also relax, start by relaxing these modes and reducing the total current. It is not until $d/l_{ee} > 1 $, as in Fig.~\ref{fig:Distribution}(e), that they start improving the conductance. On the contrary, tomographic dynamics preserve the odd parity modes, so the accumulation of electrons parallel to the channel is still visible at high values of $d / l_{ee}^{\rm even}$. 

\section{Navier-Stokes equation}

\begin{figure}[ht]
    \centering
    \includegraphics[width=0.65\linewidth]{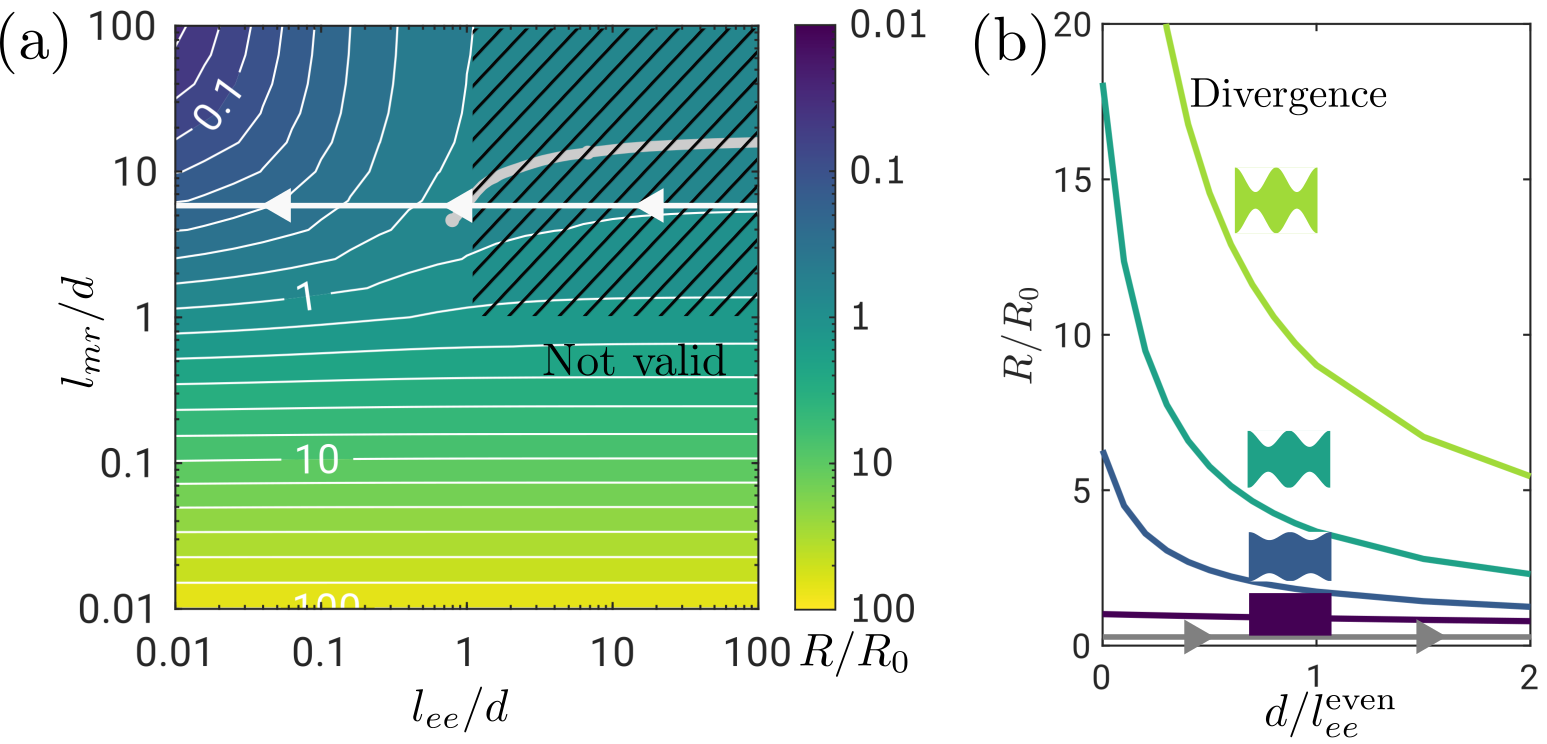}
    \caption{The Navier-Stokes equation is not the solution to the superballistic paradox. (a) Resistance is a function of $l_{mr}$ and $l_{ee}$, not distinguishing whether the odd parity modes relax. Not that the Navier-Stokes model is not valid in the striped area~\cite{alternative_routes_to_electron_hydrodynamics}. (b) Resistance as a function of $l_{ee}$ for uniform and crenelated channels at $l_{mr} / d = 5$.}
    \label{fig:navier}
\end{figure}

The superballistic paradox may be overlooked if we use standard hydrodynamic models for traditional fluids that do not distinguish classical and tomographic dynamics. Let us show, however, that this hydrodynamic models are not the solution. 
If electron transport is collective~\cite{alternative_routes_to_electron_hydrodynamics}, for example, when $l_{ee} \ll d$, it is possible to derive a hydrodynamic model from the Boltzmann transport equation. The drift velocity $\bm u$ is given by the continuity equation 
\begin{equation}
\nabla \cdot \bm u = 0  \, ,
\end{equation}
and the Navier-Stokes equation
\begin{equation}
-\nu \nabla^2 \bm u  + \frac{v_F}{l_e} \bm u = \frac{e}{m} \nabla V \, , 
\end{equation}
where $ \nu = v_F/ \left[4 \left( 1/l_{mr} + 1/l_{ee} \right)\right]$ is the viscosity. Edge scattering is modeled by a slip length given by 
\begin{equation}
\xi = \frac{3\pi }{4} \frac{\nu }{v_F}\ ,
\label{slipLength}
\end{equation}
for diffusive edge scattering, and $\xi \to \infty$ for specular scattering. The Navier-Stokes model can be solved analytically in a uniform channel 
\begin{equation}
R = \frac{R_0 d }{l_{mr}}\left[ 1 -  \frac{2 D_\nu /d }{\coth (d/2D_\nu)  + \xi/D_\nu}\right]^{-1} \, , 
\end{equation}
where $D_\nu = \sqrt{\nu l_{mr}  / v_F}$, and numerically in crenelated channel. However, this model does not distinguish the even and odd parity $l_{ee}$, and is not valid in the ballistic regime. Figure~\ref{fig:navier} shows the results we would obtain if we tried to study the superballistic paradox with the Navier-Stokes model. Furthermore, using the Navier-Stokes equation would result in another paradox due to the huge increase, nearly a divergence, in the resistance at $d/l_{ee} \to 0 $. The subsequent percentages of decrease in the resistance would be inconsistent with those obtained in experiments $\lesssim 20 \, \%$~\cite{superballistic_flow_of_viscous_electron_fluid_through_graphene_constrictions,superballistic_conduction_in_hydrodynamic_antidot_graphene_superlattices}. Alternatives to the Navier-Stokes equation, such as the anti-Mathiessen rule, have been derived for a point-contact~\cite{higher_than_ballistic_conduction_of_viscous_electron_flows,superballistic_flow_of_viscous_electron_fluid_through_graphene_constrictions}. However, it does not explicitly distinguish each type of dynamics as it poses when generalized for a magnetic field~\cite{superballistic_conduction_in_hydrodynamic_antidot_graphene_superlattices}. Non-local conductivity tensors have also been used in specific geometries~\cite{tomographic_dynamics_and_scale_dependent_viscosity_in_2D_electron_systems,linear_in_temperature_conductance_in_electron_hydrodynamics}. In this work, we solve the Boltzmann equation directly to describe not only hydrodynamic but also ballistic flow. 

\section{Collision integrals}

Let us explore the microscopic origin of tomographic dynamics, solving the collision integral that gives the electron-electron scattering $\Gamma_{ee}$. Let $f(\bm k)$ be the electron distribution function depending on $k$ and $\theta$. Therefore, under the linear approach, the rate of change in $f$ at $\bm k_1$ is given by the following integral~\cite{anomalously_long_lifetimes_in_two_dimensional_fermi_liquids,tomographic_dynamics_and_scale_dependent_viscosity_in_2D_electron_systems,the_hierarchy_of_excitation_lifetimes_in_two_dimensional_fermi_gases}
\begin{align}
\Gamma_{ee} [ f(\bm k_1 )  ] = &-\frac{2\pi }{\hbar} \iiint
\delta \left( \varepsilon (\bm k_1) + \varepsilon (\bm k_2) - \varepsilon (\bm k_3) -\varepsilon (\bm k_4) \right) \, \delta^2 \left( \bm k_1 + \bm k_2 - \bm k_3 - \bm k_4 \right) \, \left\lvert \langle \bm k_1 \bm k_2 | \mathcal{H} | \bm k_3 \bm k_4 \rangle \right\rvert^2  \, {4\pi^2} \, \nonumber \\
\times & \left[f(\bm k_1) f(\bm k_2 ) \left( 1- f(\bm k_3 ) \right)  \left( 1- f(\bm k_4 ) \right) -  \left( 1- f(\bm k_1 ) \right)  \left( 1- f(\bm k_2 ) \right) f(\bm k_3) f(\bm k_4 ) \right] \, \frac{{\rm d}^2 \bm k_2 }{4\pi^2}\, \frac{ {\rm d}^2 \bm k_3}{4\pi^2} \, \frac{{\rm d}^2 \bm k_4}{4\pi^2}\ , 
\end{align}
%
%{\color{red} No se dice qu\'{e} es $\mathcal{H}$. ¿Hay apantallamiento o es Coulomb $\sim 1/k^2$?}\\
%{\color{red} ¿De d\'{o}nde sale el factor $4\pi^2$} tras el elemento de matrix? No lo veo si considero la regla de oro de Fermi.\\ \\
%
which counts how many electrons with wavevector $\bm k_1 $ collide with other electrons with momentum $\bm k_2$ to end up with momenta $\bm k_3$ and $\bm k_4$. The delta functions account for energy and momentum conservation in the scattering process. The probability is weighted, considering the occupation of the initial and final states. The term with a minus also considers the inverse collisions that bring electrons from arbitrary states $\bm k_3$ and $\bm k_4$ back to $\bm k_1$ and $\bm k_2$. We consider an interaction Hamiltonian, $\mathcal{H}$, such that the scattering matrix element is constant $V = | \langle \bm k_1 \bm k_2 | \mathcal{H} | \bm k_3 \bm k_4 \rangle|$, showing that tomographic dynamics arises from the conservation of momentum and energy, together with the Fermi-Dirac distribution. Transport is characterized by the Fermi wavenumber $k_F$ and velocity $v_F$. Last, we assume a linear band dispersion $\varepsilon (\bm k) = \hbar v_F k$. The collision operator reads 
\begin{align}
\Gamma_{ee} [ f(\bm k_1 )  ] = & - \frac{ V^2}{8 \pi^3 v_F \hbar^2 } \iiint
\delta \left( k_1 + k_2 - k_3 -k_4 \right) \, \delta^2 \left( \bm k_1 + \bm k_2 - \bm k_3 - \bm k_4 \right)   \nonumber \\
\times & \left[f(\bm k_1) f(\bm k_2 ) \left( 1- f(\bm k_3 ) \right)  \left( 1- f(\bm k_4 ) \right) -  \left( 1- f(\bm k_1 ) \right)  \left( 1- f(\bm k_2 ) \right) f(\bm k_3) f(\bm k_4 ) \right] \, {{\rm d}^2 \bm k_2 }\, { {\rm d}^2 \bm k_3} \, {{\rm d}^2 \bm k_4}\ .
\end{align}

First, we study a thermal distribution such that 
$f( \bm r, \bm k)$ is given by the Fermi-Dirac distribution, with a small deviation $\delta f$, and let us define
\begin{equation}
\psi (\bm k ) = \frac{\delta f(\bm k)}{f(\bm k) \,\left( 1 - f(\bm k) \right)}\ .
\end{equation} 

Following the approach in~\cite{anomalously_long_lifetimes_in_two_dimensional_fermi_liquids}, let us rewrite a collision operator for $\psi$, as 
\begin{equation}
\Gamma_{ee} [\psi (\bm k ) ] = - \frac{ k_F V^2}{4\pi v_F \hbar^2 } \iint f(\bm k_1) f(\bm k_2) \left( 1- f(\bm k_3) \right) \left( 1- f(\bm k_4) \right) \left( \psi (\bm k_1 )  + \psi (\bm k_1 )   - \psi (\bm k_2 )  -\psi (\bm k_4 ) \right) {\rm d \theta^* } \, {\rm d} \bm k_2 \ ,
\end{equation}
where we simplify some integrals using the properties of the delta function and obtain a numerically solvable problem in terms of $\theta$, the angle between the final and initial momenta. Therefore, using energy and momentum conservation, for each set of $\bm k_1$, $\bm k_2$ and the integration parameter $\theta^*$ corresponding to the angle between the initial and final states, we can solve the final $\bm k_3$ and $\bm k_4$. Still, this expression is not implemented directly in the Boltzmann transport equation. Instead, we write $\psi$ as a sum of eigenmodes $\psi_m (\bm k ) $ such that 
\begin{equation}
\Gamma_{ee} [\psi (\bm k ) ] = -\frac{1}{\tau_{ee}^{(m)}} \, \psi (\bm k)\ , 
\end{equation}
where $\tau^{(m)}$ is the relaxation time and $l_{ee}^{(m)} = v_F \tau^{(m)}$ is the relaxation length for a particular mode. Still, this formalism depends on $\bm k = (k,\theta)$, but the adapted Boltzmann equation, which considers that transport phenomena occur near the Fermi surface, only depends on $\theta$. Let us neglect the $k$ dependence and consider the set of orthogonal modes $\psi^{(m)}(\bm k ) = \cos m \theta$ or $\sin m \theta$, where $m = 1, 2 , 3 \dots$. Given the rotational symmetry of the problem and the equilibrium function, they can be used a set of eigenvectors, and the eigenvalues are given by 
\begin{equation}
\frac{1}{\tau_{ee}^{(m)}} = \frac{1}{v_F l_{ee}^{(m)}}= -\,\frac{1}{\pi } \int_{0}^{2\pi} \psi^{(m)} (\bm k ) \, \Gamma_{ee} [\psi^{(m)} (\bm k ) ] \, {\rm d} \theta \, . 
\end{equation}

It is immediate to transform the Boltzmann equation in terms of $f ( \bm r, \bm k)$ to the equation in terms $g(\bm r, \theta ) $, that can be solved for a given geometry. In this work, we shall further simplify the number of parameters in the Boltzmann equation and use a single $l_{ee}^{\rm even}  = l_{ee}^{(2)} $ representative of all the even parity modes, and a single $l_{ee}^{\rm odd} = l_{ee}^{(3)}$ representative of all the odd parity modes, since $m = 3$ is the mode that contributes the most to electron transport. Odd parity modes go as $l_{ee}^{\rm (m) } \sim m^4 (T_F/T)^2 l_{ee}^{\rm even}$ at low temperatures, so, to study the initial behavior of the curves, we can follow this approach. We could increase the number of terms $m=2, 3, 4, 5 \dots$, but that would require either (i) including more parameters in the calculation, making it hard to understand whether experimental results correspond to a hydrodynamic or tomographic description or (ii) to write the mean free paths as a function of $T/T_F$, where $T_F$ is the Fermi temperature. However, if we do that, these collision integrals will be part of the demonstration of tomographic behavior. Following our approach, the tomographic dynamics is proved without any assumption on the functional dependence of $l_{ee}^{\rm even}$ and $l_{ee}^{\rm odd}$. Furthermore, our results show that two parameters are enough to solve the superballistic paradox. \\

\begin{figure}[ht!]
    \centering
    \includegraphics[width=0.5\linewidth]{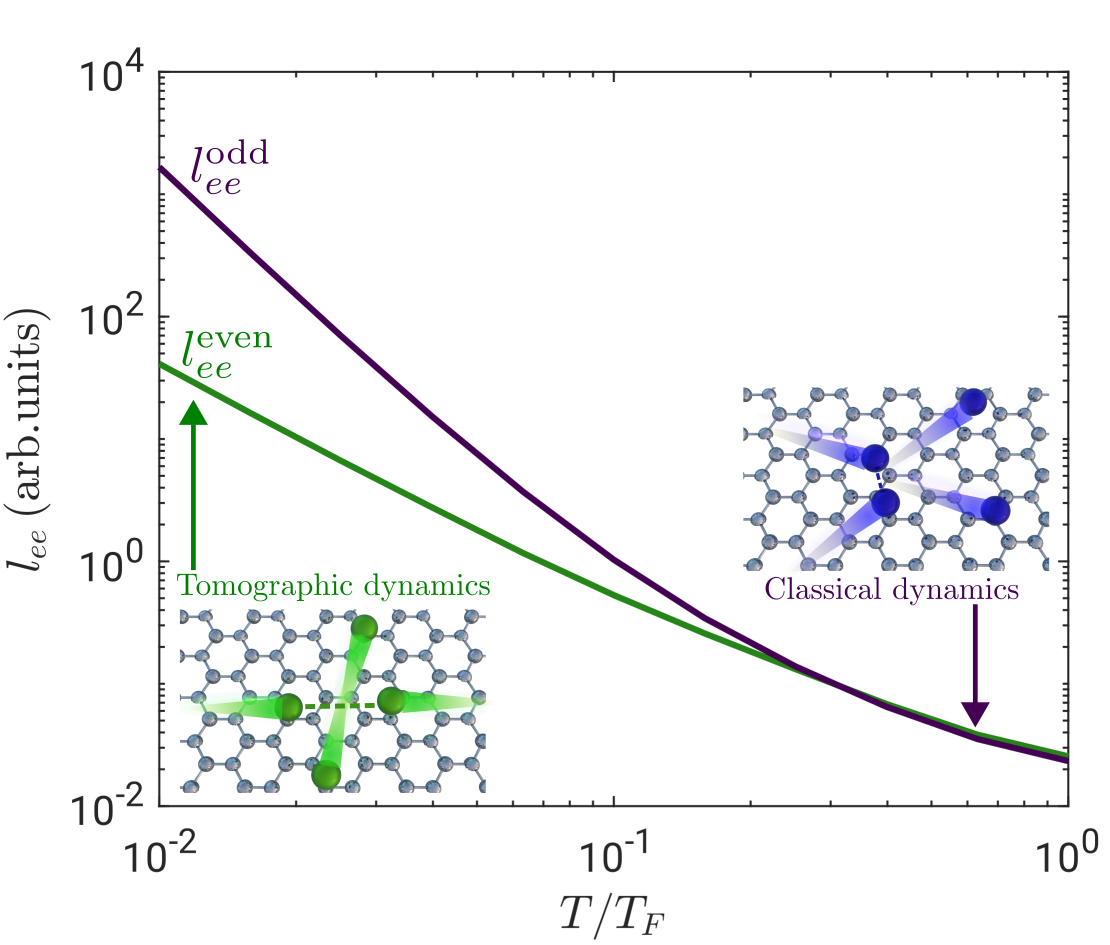}
    \caption{Mean free paths $l_{ee}^{\rm even} \simeq l_{ee}^{(2)}$ and  $l_{ee}^{\rm oddd} \simeq l_{ee}^{(3)}$ as a function of the ratio between the temperature and the Fermi temperature. Dynamics are tomographic at low temperatures $T \ll T_F$, where $l_{ee}^{\rm odd} \gg l_{ee}^{\rm even}$, while they converge to classical dynamics at $T \sim T_F$.   }
    \label{fig:CollisionIntegrals}
\end{figure}

Figure~\ref{fig:CollisionIntegrals} shows the numerical results for the mean free paths $l_{ee}^{\rm even}$ and $l_{ee}^{\rm odd}$. In graphene, $T_F \simeq 1000 \, \rm K$ at $n = 0.5 \times 10^{12} \, \rm cm^{-2}$, so it is only at high temperatures that $l_{ee}^{\rm even}$ and $l_{ee}^{\rm odd} $ are comparable. Low-temperature dynamics are tomographic $l_{ee}^{\rm odd} \gg l_{ee}^{\rm even} $, resulting in a decreasing resistance as observed in experiments and our simulations of the Boltzmann transport equation. 

\section{Non-thermal collision integrals}

The former calculations include the Fermi distribution as a key ingredient in the derivation of tomographic dynamics. Indeed, we can only ensure that head-on collisions are predominant when the distribution is close to a sharp step. Here, we introduce a non-thermal distribution of electrons, similar to the Fermi distribution
\begin{equation}
    f(\bm k ) = \frac{1}{1+\exp \left[  \frac{ \hbar v_F  k - {\rm sign}(g(\theta) )  k_B T(\theta) }{  k_B T(\theta)} \right ] }
    \label{fermiNonThermal}
\end{equation}
where the width of the distribution $|T(\theta)| $ depends on the angle, and is given by Eq.~\eqref{gAnayltical} as
\begin{equation}
    T(\theta ) = \frac{\bar{T}}{\frac{1}{2\pi}\int_{0}^{2\pi } |g(\theta ) | \, {\rm d} \theta}\, | g(\theta)  | 
    \label{Tproptog}
\end{equation}
meaning that the distribution is wider in the directions and is parallel to the axis of the channel, where the electrons are brought out of equilibrium by a high current and narrower in the other directions. This approach, valid in the absence of electron-electron collisions, explains whether resistance increases or decreases in the low-temperature limit. We also consider the shift of the distribution in the direction of the current with the term ${\rm sign}(g(\theta)) k_B T(\theta) $. However, the particular shape of the distribution does not significantly alter the conclusions, provided that it is a function that is wider in the direction of the channel. We can no longer define a temperature, but for comparison, we describe our results as a function of an average $\bar{T}$ found upon integration over all the angles. \\

If we take an arbitrary electron, it would be more likely to overcome the Pauli blockade by colliding against an electron that moves parallel to the channel, where the distribution is widest. The sharper the distribution function, the more these new types of electron-electron collisions will occur. Suppose the distribution is sharp enough for a long value of $l_{\rm mr} / d$, see Eq.~\ref{gAnayltical}. In that case, these new collisions will dominate the tomographic head-on ones, and the odd-parity modes could relax via collisions with the electrons parallel to the channel. \\

\begin{figure}[ht!]
    \centering
    \includegraphics[width=\linewidth]{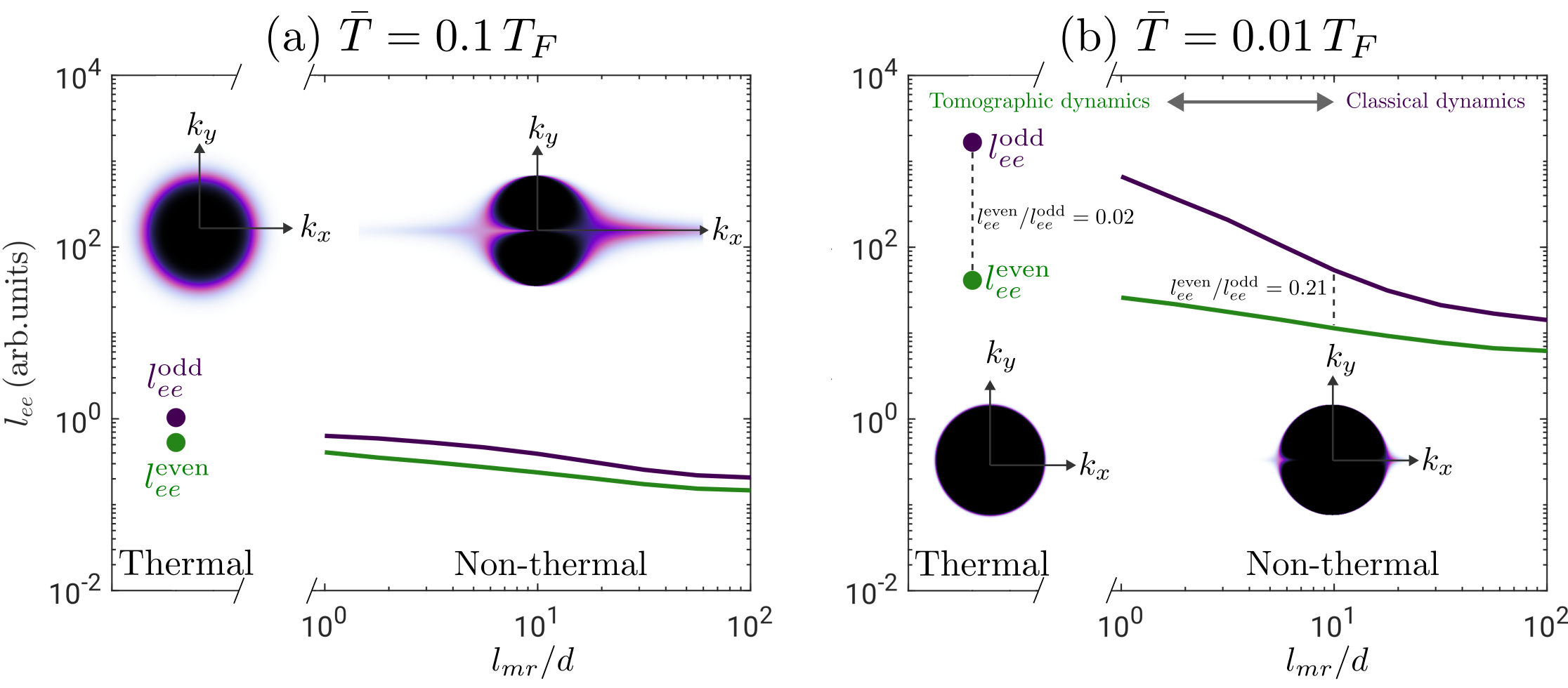}
    \caption{
    %Mean free paths $l_{ee}^{\rm even}$ and  $l_{ee}^{\rm oddd}$ for thermal and non-thermal distributions, as a function of the mean free path $l_{mr}$ against defects. (a) Equivalent temperature $\bar{T} = 0.1 \, T_F$, we show the electron distributions. (b)~Equivalent temperature is $\bar{T} = 0.01 \, T_F$. Non-thermal distributions with low collisions favor transitioning from tomographic to dynamics closer to the classical ones. 
    Mean free paths $l_{ee}^{\rm even}$ and  $l_{ee}^{\rm odd}$ for thermal (solid dots) and non-thermal (solid line) distributions, as a function of the mean free path $l_{mr}$ against defects. Electron distributions and results obtained for equivalent temperature (a) $\bar{T} = 0.1 \, T_F$ and (b) $\bar{T} = 0.01 \, T_F$. Reducing collisions against defects favors transitioning from tomographic dynamics towards dynamics closer to the classical ones.}
    \label{fig:CollisionIntegralsNonThermal}
\end{figure}

Let us show some additional calculations of the relaxation times using the non-thermal distribution in Eq.~\eqref{fermiNonThermal}. As the problem does not have polar symmetry, the angular harmonics are no longer the eigenmodes of the collision operator, so any polar mode is allowed to decay to other modes partially. However, the off-diagonal coefficients are smaller than the diagonal ones, so we focus on computing the diagonal terms to discuss the electronic properties. In particular, we find $l_{ee}^{\rm even}$ for the $\cos (2 \theta) $ mode, which is similar to the $\sin (2 \theta) $ value, and $l_{ee}^{\rm odd} $ for the $\cos (3 \theta)$ mode. 
Figure~\ref{fig:CollisionIntegralsNonThermal}(a) shows the mean free paths for thermal and non-thermal distribution functions. For the thermal distribution, we show $l_{ee}^{\rm even} $ and $l_{ee}^{\rm odd}$ that does not depend on the particular value of $l_{mr}/d$. In the case of the non-thermal distribution, the represented values are $l_{ee}^{\rm even(m=2)} $ and $l_{ee}^{\rm odd(m=3)}$ estimated for $m=2$ and $M=3$ respectively. To numerically evaluate the latter for different values of $l_{mr}/d$, we need to set a minimum $T_{\rm min}$ in all directions, replacing $T(\theta ) \to T(\theta ) + T_{\rm min} $ in  Eq.~\ref{Tproptog}. Then, we compute the limit when this parameter goes to zero by setting $T_{\rm min} = 0.0001 \, T_F$, and compute the integrals with a relative tolerance of $1\%$. \\

%Figure~\ref{fig:CollisionIntegralsNonThermal}(a) shows the mean free paths for thermal and non-thermal distribution functions. 
At low temperatures, as shown in Fig.~\ref{fig:CollisionIntegralsNonThermal}, the dynamics of thermal distributions are tomographic, with $l_{ee}^{\rm odd} \gg l_{ee}^{\rm even}$. On the contrary, the even and odd mean free paths become comparable for non-thermal distributions with increasing $l_{mr} / d$. According to previous results in Fig.~\ref{fig:Cuts}, even the ratio $l_{ee}^{\rm even} / l_{ee}^{\rm odd} = 0.21$ obtained for $l_{\rm mr} / d = 10$ is closer to classical dynamics than to tomographic dynamics, exhibiting an initial increase in the resistance. 
%Despite an accurate calculation would require computing a general collision operator $\Gamma_{ee} [g]$ with off-diagonal terms, 
Within our assumptions, our calculations suggest that tomographic dynamics, which are highly dependent on the Fermi distribution, do not apply to non-thermal distributions, highlighting a key particularity of Molenkamp's experiment.

\end{document}